\documentclass[
aps,
prd,
superscriptaddress,
nofootinbib,
floatfix]
{revtex4}
\usepackage{graphicx,
longtable}
\usepackage{amsmath,amstext,amsfonts,amsbsy,amssymb,amscd,bbm,epsfig,lscape}

\begin{document}

\title{A scalar field coupled to a brane in ${\cal M}_4 \times {\cal S}_1$. Part I: Kaluza-Klein spectrum and zero-mode localization}

\begin{flushright}
IFIC/15-90
\end{flushright}

\author{A.~Donini}
\affiliation{
Instituto de F\'{\i}sica Corpuscular, CSIC-Universitat de Val\`encia,\\
       Apartado de Correos 22085, E-46071 Valencia, Spain
}

\begin{abstract}
\vspace*{1cm}
A toy model where a massless, real, scalar field $\Phi$ in a compact space-time ${\cal M}_4 \times {\cal S}_1$ is coupled to a brane (parametrized as a $\delta$-function) through
the unique relevant operator $\delta (y) \Phi^2 (x,y)$ is considered. 
The exact Kaluza-Klein spectrum of the model is computed for any value of the coupling between field and brane using the Burniston-Siewert method
to solve analytically transcendental equations. The exact KK-spectrum of a model with a Brane-Localized Kinetic Term is also computed.
Weak- and strong-coupling limits are derived, matching or extending mathematically equivalent existing results. 
For a negative coupling, the would-be zero-mode $\psi_{0^-}^e$ is found to localize into the brane, behaving as an effective four-dimensional field.
The 4-dimensional KK-decomposition of the model once a renormalizable cubic self-interaction $\Phi^3 (x,y)$ is added to the action
is derived computing the overlaps between the KK-modes. It is found that the localized would-be zero-mode $\psi_{0^-}^e$ decouples from
the massive KK-spectrum in the limit of large brane-to-bulk coupling.
\end{abstract}
\medskip
\pacs{
}

\maketitle

\section{Introduction}
\label{sec:intro}

The recent discovery of a scalar particle with a mass $m^{\rm exp}_H = 125.7 \pm 0.4$ GeV \cite{Agashe:2014kda} in 2012 by 
the ATLAS and CMS Collaborations (see Refs. \cite{Aad:2013wqa,Aad:2014aba,Aad:2014eva,Aad:2014eha}  and 
\cite{Chatrchyan:2013iaa,Chatrchyan:2013mxa,Khachatryan:2014ira} for recent results), together with the (current) absolute
lack of any evidence of physics beyond the Standard Model, dramatically points out our poor theoretical understanding of the physics governing electroweak
symmetry breaking. Is the Standard Model of Fundamental Interactions a renormalizable theory? What determines the scale of the symmetry breaking and the
Higgs mass if no new physics permit to relate it with some, more fundamental, process? 
With stubborn determination, we prefer to keep on believing that the Standard Model is not the end of the story. This hypothesis is justifiable for several 
theoretical and experimental reasons. First of all, the Standard Model cannot explain the observed dark matter component of the Universe energy density, 
$\Omega_{\rm DM} \sim 27$\%;  it  has no clue for the so-called {\em dark energy}  that should determine the observed accelerated expansion of the 
Universe, $\Omega_{\rm DE} \sim 0.68$\%;  the amount of CP violation in the Standard Model is not enough to explain Baryogenesis; and, eventually, 
the observation of non-vanishing neutrino masses cries for an extension of the Standard Model that could account for them.
In addition to these experimental hints of physics beyond the Standard Model, theorists  have a strong prejudice in favour of 
Wilsonian effective theories: a given model can only explain phenomena up to a certain scale through a finite set of relevant operators, above which scale 
the model should be replaced by a new one that would incorporate new degrees of freedom and will explain new classes of phenomena\footnote{
Notice that this approach is in open contrast with the accepted paradigm of mid-XX century, for which only renormalizable field theories would make sense.}.
If the Standard Model were a low-energy effective theory, however, a theoretical inconsistency would be implied by $m_H^{\rm exp} = {\cal O} (\Lambda_{\rm EW})$, with $\Lambda_{\rm EW} \sim 246$ GeV the electro-weak symmetry breaking scale. As the scalar mass radiative corrections are quadratic in the cut-off
(differently from fermion masses), in an effective theory one would naively expect that a scalar particle mass would be sensitive (quadratically) to the scale at which the theory
does not make sense any longer. If that scale is the Planck scale, $M_P \sim 10^{19}$, then  {\em naturalness}  \cite{'tHooft:1979bh} implies that $m_H^{\rm th} = {\cal O} (M_P)$. 
The large hierarchy between $m_H^{\rm exp}$ and the expected value $m_H^{\rm th}$ is called the {\em hierarchy problem}, and is the most important theoretical motivation 
for the existence of new physics (much) below $M_P$. This line of thought started (a little before and a little after 1980) both the rise of Technicolor \cite{Weinberg:1975gm,Weinberg:1979bn,Susskind:1978ms}  in its various manifestations and of the Minimal Super-Symmetric Standard Model \cite{Dimopoulos:1981zb}.

In the '90s,  yet another proposal was advanced to solve the hierarchy problem \cite{Antoniadis:1990ew,Antoniadis:1997zg,ArkaniHamed:1998rs,Antoniadis:1998ig}. 
If we want to explain the large hierarchy between $\Lambda_{\rm EW}$ and $M_P$ without introducing new physics in between, why don't we lower $M_P$, instead? 
This could be done assuming the existence of new spatial dimensions in excess of the observed three ones to which we are used to at human-being length scales. 
In order for these new dimensions to pass unnoticed to the eye of an observer, they must be compactified in such tiny volumes that direct observation through the measurement of
deviations from the inverse-square Newton's law for gravitational interactions is beyond the reach of current experiments \cite{Adelberger:2009zz}.  
Present limits on new spatial dimensions gives $R \leq 44 \, \mu$m at 95\% CL for the largest extra-dimension compactified in a circle of radius $R$ \cite{Kapner:2006si}. 
Imagine now that gravity may propagate into this tiny compact volume (called {\em bulk}) at very small distances, $r \ll R$. For these scales, gravity is $D$-dimensional 
and its fundamental scale is $M_D$.  On the other hand, at distances $r \gg R$ the gravitational potential behaves as effectively 4-dimensional. Matching the long distance
limit of the $D$-dimensional gravitational potential and the Newtonian 4-dimensional one \cite{Kehagias:1999my,Floratos:1999bv}, we get the relation
\begin{equation}
M_P^2 \propto V_n \times M_D^{2 + n}
\end{equation}
being $n$ the number of extra spatial dimensions and $V_n$ the volume of the compact dimensions, {\em i.e.} for a $n$-torus $V_n = (2 \pi R)^n$.
This relation was first obtained in Refs.~\cite{Antoniadis:1997zg,ArkaniHamed:1998rs,ArkaniHamed:1998nn} and it states that, if $V_n$ is large enough, the fundamental mass scale $M_D$ can actually be much lower than $M_P$ and, possibly, as low as the electro-weak symmetry breaking scale $\Lambda_{\rm EW}$, thus solving the hierarchy problem. For $n = 1$,  the radius $R$ must be of astronomical size to have $M_D \sim 1$ TeV. However, for $n \geq 2$ to lower $M_D$ down to some TeV's a sub-mm radius $R$ suffices, something that is not excluded by direct observation of deviations from the Newtonian 4-dimensional gravitational law, as stated above.
Being a large compact volume the origin of a large 4-dimensional Planck mass, this solution to the hierarchy problem is called {\em Large Extra-Dimensions} (LED).

The idea of compact dimensions is not at all new. It was proposed by T.~Kaluza in 1919 with the intent of unifying electromagnetism and gravity at the classical level~\cite{Kaluza:1921tu}. It was later extended to include quantum mechanical concepts by O.~Klein in 1926 \cite{Klein:1926tv} (albeit a complete derivation of the equations of motions for 
$G_{MN}$, with $M,N = 0,\dots,3+n$ was completed only in the '40~\cite{Scherrer:1941,Jordan:1946,Lichnerowicz:1947,Jordan:1947,Ludwig:1947,Thiry:1948a,Thiry:1948b,Jordan:1948,Ludwig:1948,Scherrer:1949a,Scherrer:1949b}). It was thus shown that, at the classical level, it is possible to understand electromagnetism as a phenomenon related to the existence of compact extra-dimensions by identifying the photon field $A_\mu$ with $G_{\mu 5}$\footnote{In addition to $G_{\mu\nu}$ and $G_{\mu 5}$, a new scalar degree of freedom should be included, $G_{55}$, eventually leading to the development of the Brans-Dicke extension of 4-dimensional general relativity \cite{Brans:1961sx}.}. 
With the discovery of nuclear forces, the idea of unifying 4-dimensional gravity with elementary particles interactions was pursued (see, for example, the classical papers~\cite{Salam:1981xd,RandjbarDaemi:1983jz}). It suffices here to say that unification of gravity with all of the SM gauge interactions into a unique multi-dimensional metric is not a simple task and that it has not been possible up to now to find a simple, elegant, solution to this problem as it was possible with the unification of gravity and classical electromagnetism, as seen above. When the extra-dimensional idea is applied to solve the hierarchy problem, then, SM gauge fields are added as new, multi-dimensional, degrees of freedom ($A_M,W^i_M$ and $A^a_M$ representing the photon, the intermediate vector bosons and the gluons, respectively). 

This approach introduces a new problem, though. A straightforward consequence of compactification is that the momentum $q$ of a field in the compact dimension gets quantized.
As an example, consider a field $\Phi (x_\mu,y)$, being $x_\mu$ the standard four dimensions and $y$ an extra dimension compactified on a circle of radius $R$, such that $\Phi (x_\mu, y + 2 \pi R) = \Phi (x_\mu,y)$. The field $\Phi (x_\mu,y)$ can be decomposed into an infinite tower of 4-dimensional fields $\phi_n (x_\mu)$:
\begin{equation}
\label{eq:KKdecomposition}
\Phi (x_\mu,y) = \sum_{n = -\infty}^\infty c_n \phi_n (x_\mu) e^{i n y/ R} \qquad {\rm with \, energy} \qquad E = \sqrt{m^2 + |\vec {p}|^2 + \left ( \frac{n}{R} \right )^2} \, ,
\end{equation}
with $m$ the mass of the 5-dimensional field $\Phi (x_\mu,y)$, $\vec{p}$ the 3-dimensional spatial momentum and $c_n$ normalization coefficients. Notice that the dispersion relation in eq.~(\ref{eq:KKdecomposition}) states that the 4-dimensional fields $\phi_n (x_\mu)$ have 4-dimensional masses $m_n = \sqrt{m^2 + (n/R)^2}$. This infinite tower of massive fields is called {\em Kaluza-Klein tower}.
If the photon field $A_M (x_\mu,y)$ is compactified on a circle, then, infinite photons with masses $m_n = n/R$ should be observed experimentally. For $R \simeq 1$ $\mu$m this corresponds to a mass $m_n \simeq (0.2 \times n)$ eV, in contrast with present bounds on $m_\gamma$.  A solution to this problem was given in Refs.~\cite{Antoniadis:1997zg,ArkaniHamed:1998rs,Antoniadis:1998ig} by introducing the concept of {\em branes} 
in the realm of particle phenomenology. The concept was not new by the time, but was only applied in the context of string theory \cite{Polchinski:1996na}, 
being actually the driving force beyond what is called the {\em second string revolution}. A $p$-brane is a $p$-dimensional object embedded in a space-time with $n \geq p$ spatial dimensions. A $Dp$-brane is a topological defect with $p$ spatial dimensions defined as the locus in space-time where the ending points of an open string are bounded by
Dirichlet boundary conditions. In the context of string theory, low-energy excitations of the endings of an open string represent gauge fields. If those endings are fixed to some
$p$-dimensional object, thus, gauge fields get localised on it and  cannot span the spatial dimensions transverse to the brane. After compactification of those dimensions, thus, 
no Kaluza-Klein towers will arise in correspondance to the gauge degrees of freedom. The model of Refs.~\cite{Antoniadis:1997zg,ArkaniHamed:1998rs,ArkaniHamed:1998nn}
is built, then, around these two concepts: gravity permeates a $(4+n)$-dimensional space-time, of which $n$ spatial dimensions are compactified in a relatively large volume
$V_n$ (to solve the hierarchy problem), whereas the SM fields are localised on a non-compact D3-brane with 4-dimensional Minkowski metric 
(see, however, Refs.~\cite{Donini:1999px,Antoniadis:1999bq,Accomando:1999sj} for models in which SM fields have KK-excitations, too, 
yet still solving the hierarchy problem). 
The model contains, thus, the Standard Model particle content plus an infinite tower of Kaluza-Klein gravitons with masses $|\vec{n}|/R$.

The implementation of the brane concept at the level of field theory is not unique and several different approaches  have been pursued. 
An action principle was introduced in Ref.~\cite{Sundrum:1998sj} based on the following idea: the action contains two terms, 
a term corresponding to fields that can see the whole $(4+n)$-dimensional space-time ${\cal M}_{4+n}$, and a second term containing fields explicitly
stuck on the 4-dimensional brane ${\cal M}_4$ (the Standard Model fields). Assuming a flat geometry, for simplicity:
\begin{equation}
S = S_{\rm bulk} + S_{\rm brane} = \int d^4 x \int_{V_n} d^n y \, {\cal L }[\Phi_{\rm bulk} (\vec x, \vec y)] 
+ \int d^4 x  \int_{V_n} d^n y \, \delta^{(n)}(\vec{y}) \{ - f^4 + {\cal L} [\phi_{\rm brane}(\vec x, \vec y)] \} \,  ,
\end{equation}
being $\Phi_{\rm bulk}$ a generic bulk field, to be decomposed into a tower of 4-dimensional fields according to eq.~(\ref{eq:KKdecomposition}),
$\phi_{\rm brane}$ a generic brane field and $f$ the brane tension (the constant term proportional to $f^4$ represents the brane contribution to the cosmological constant). 
The most important consequence of this approach is that the brane, located here at $\vec y = \vec 0$, breaks explicitly translational invariance in the extra-dimensions and, thus, 
extra-dimensional momenta are not conserved\footnote{In the absence of a potential in the extra-dimensions to fix the brane position at $\vec y = \vec 0$, the mere presence of the brane still breaks translational invariance in the extra-dimensions, albeit spontaneously. Goldstone bosons related to this symmetry breaking are the coordinates of the
brane themselves, $Y_i(\vec x)$ (see Ref.~\cite{Sundrum:1998sj}).}. Once gravity is turned on, thus, KK-gravitons may be emitted in the bulk by SM particles up to the highest kinematically allowed KK-number with a universal coupling, giving interesting experimental signals (see Ref.~\cite{Giudice:1998ck}). Partial conservation of momentum in the extra-dimensions can be obtained replacing 
the $\delta^{(n)}(\vec y)$ function fixing the brane position in the bulk by a form factor $B (\vec y)$ defining a certain shape of the brane in the extra-dimensions 
related to its tension $f$~\cite{DeRujula:2000he}. Brane fields stuck onto this {\em fat brane} are no longer insensitive to extra-dimensions and have their own Kaluza-Klein towers related to the brane size $L$  ($L \sim 1/f$) and not to the compactification radius $R$ (see Refs.~\cite{Macesanu:2002db,Macesanu:2002ew} for experimental signatures characteristic to this model). Notice that, in the absence of gravity, this approach is commonly known as {\em Universal Extra Dimensions} (UED) \cite{Appelquist:2000nn} in the literature. 
Its motivation is that momentum in the extra-dimensions is conserved and, thus,  the lightest of the KK-excitations of the SM particles is guaranteed to be stable, being a possible dark matter candidate (see Ref.~\cite{Servant:2014lqa} and references therein for the present status of UED models).

In this paper I will follow a slightly different approach. I restrict myself to the simple case of a scalar field $\Phi(x,y)$ living in a 5-dimensional space-time with metric 
${\cal M}_4 \times {\cal S}_1$. Notice that, as only one extra spatial dimension is considered, this model cannot solve the hierarchy problem (at least two extra-dimensions are needed to that purpose). The {\em brane} is then introduced as a topological defect in the extra-dimensions: it is not the mathematical locus where SM fields are bounded to stay, but an external object living in the same space-time where bulk fields live, with its own physical properties (its own tension, certainly, but also a momentum or an angular momentum, possibly). 
Considering the brane a (solitonic) object in the extra-dimension is not new: see Ref.~\cite{Csaki:2004ay} and refs. therein for some example. 
Notice that  it is not the purpose of this paper to explain the origin of the defect, whose origin may be possibly ascribed to the underlying quantum structure of the gravitational vacuum (see, {\em e.g.},  Ref.~\cite{Hawking:1976jb}). 
If only one extra-dimension is considered (the case discussed in this paper), the defect can be represented by a $\delta$-function\footnote{In higher dimensions, $\delta$-functions are still an option, but more complicated defects may be imagined. For example, in Refs.~\cite{Frere:2000dc,Frere:2001ug,Frere:2003yv,Frere:2010ah} the {\em brane} is represented by a vortex in two extra-dimensions.}. This picture is rather intuitive from the visual point of view:  the {\em brane} is an infinitely thin object localized in the extra-dimension, {\em i.e.} a spike when seen by an external observer.  Once a defect is localized in the extra-dimension, bulk fields can couple to it, bumping on or passing through it depending on their momentum in the extra-dimension. The coupling of the field $\Phi$ with the defect may be described in an effective field theory approach by a tower of operators built with powers of $\Phi$, of its derivatives and of the $\delta$-function, classified by their classical dimensions. It can be shown that only one such operator is {\em relevant} in five dimensions ({\em i.e.} its coupling has positive dimension in energy), $\delta (y) \, \Phi^2 (x,y)$. I will, therefore, consider the case in which $\Phi$ is coupled  {\em minimally} to the brane, {\em i.e.} through the operator given above.
The effect of the spike can be mathematically computed by deriving the eigen-modes of the bulk fields coupled to the spike, {\em i.e.} its
Kaluza-Klein modes\footnote{
A similar approach was adopted in the famous papers by L. Randall and R. Sundrum introducing warped extra-dimensions \cite{Randall:1999ee,Randall:1999vf}, where branes in the form of $\delta$-functions were located at fixed points in an orbifold and gravity was minimally coupled to them through the brane action cosmological constant term 
$\int d^4 x \, \int d^n y \, \sqrt{-g} \, f^4$.}. 
I  will show that even such a minimal coupling of bulk fields to the brane modifies significantly the spectrum of the theory and may induce localization of the would-be zero-mode.
This scenario could be of some interest to explain localization of SM particles onto (or, better said, into) a brane. In order to solve analytically the equation of motion of the model to derive its exact Kaluza-Klein spectrum, I will make use of a mathematical method based on complex calculus proposed some 40 years ago by Burniston and Siewert~\cite{PSP:2073132}. This method permits to solve a huge class of transcendental equations and, thus, it is particularly powerful to compute the KK-spectrum of field theories in compact space-times. For this reason, I will review in detail the method in the Appendix of this paper.
Notice that, although the initial motivation to introduce a coupling between bulk fields and the brane is different, the resulting model 
can be considered a special case of a non-minimal UED model \cite{Flacke:2013pla,Flacke:2014jwa}. In the literature on non-minimal UED, 
a special role is devoted to so-called {\em Brane-Localized Kinetic Terms} (BLKT's), whose effect is to introduce corrections to the kinetic terms of bulk fields 
at the position of the brane through (irrelevant) operators of the form $\delta (y) \times \Phi \, \Box \, \Phi + \dots$ \cite{Dvali:2001gm,Dvali:2001gx,Carena:2002me}. 
Deriving the exact spectrum of a model in which BLKT's are present is easily done using the results of the Appendix. 
After obtaining the exact KK-spectrum of the model, a self-interaction term $\Phi^3 (x,y)$ in the bulk  can be added to the action
and the corresponding self-interactions of 4-dimensional KK-modes can be easily derived by computing their overlap in the extra-dimension. 
It is of particular interest to compute the effective 4-dimensional interactions between the would-be zero-modes (both for positive and negative brane-to-bulk coupling, the latter being a localized field in the brane) and the massive KK-modes.

The paper is organized as follows:  in Sect.~\ref{sec:model} the action for a real massless scalar field $\Phi$ in a space-time ${\cal M}_4 \times {\cal S}_1$
is given together with a classification of the effective operators that may couple $\Phi$ with a brane parametrized as a $\delta$-function;  
in Sects.~\ref{sec:oddmodes} and \ref{sec:evenmodes} the KK-spectrum is computed; the main properties of the spectrum are outlined in Sect.~\ref{sec:properties}, whereas
a comparison with the results obtained in the presence of BLKT's is given in Sect.~\ref{sec:BLKT}; the weak coupling limit of the Kaluza-Klein spectrum is derived in Sect.~\ref{sec:weakcoupling} and compared with existing results (albeit on a different subject, see Ref. \cite{Dienes:1998sb}); the strong coupling limit of the Kaluza-Klein spectrum is derived in Sect.~\ref{sec:strongcoupling}; the 4-dimensional action obtained after performing a KK-decomposition of the 5-dimensional field $\Phi$ when a
 renormalizable cubic self-interaction is added to the model is computed in Sect.~\ref{sec:interactions};
I eventually conclude In Sect.~\ref{sec:concl}.


\section{Massless scalar field coupled to a brane}
\label{sec:model}

In Sect.~\ref{sec:nobrane}  I first review the simple case of a (real) massless scalar field in a 5-dimensional space-time with one spatial dimension compactified on a circle of radius $R$, with geometry ${\cal M} = {\cal M}_4 \times {\cal S}_1$ ({\em i.e.} the metric factorizes into a 4-dimensional Minkowski metric times a circle).  In Sect.~\ref{sec:withbrane} I introduce the model considered in this paper, {\em i.e.} the case in which the scalar field is coupled to a brane represented by a $\delta$-function in the extra-dimension.

\subsection{Massless scalar field in ${\cal M}_4 \times {\cal S}_1$}
\label{sec:nobrane}

Consider a free massless scalar field $\Phi(x,y)$ in a $(4+1)$-dimensional space-time, with $(3+1)$ infinitely extended dimensions 
(corresponding to our 4-dimensional space-time) with coordinates $x_\mu$, and one finite dimension labelled by the coordinate $y$, 
compactified on a circle of radius $R$, such that the action is invariant under $y \to y + 2 \pi R$. The action is just its kinetic term: 
\begin{equation}
\label{eq:scalaraction}
S = \int d^4x \int_{- \pi R}^{\pi R} dy \left \{ \frac{1}{2} \partial^M \, \Phi (x_\mu,y) \, \partial_M \, \Phi (x_\mu,y) \right \} \, .
\end{equation}
Notice that in 5 dimensions, the classical dimension of a bosonic field is 3/2 (a fermion being a dimension 2 field).

The corresponding equation of motion is: 
\begin{equation}
\label{eq:freescalar}
\Box_5 \Phi(x_\mu,y) =  \left \{ \Box_4 - \frac{d^2}{dy^2} \right \} \Phi(x_\mu,y) = 0 \, .
\end{equation}
A solution to eq.~(\ref{eq:freescalar}) can be factorized into a $4$- and an extra-dimensional component, 
$\Phi(x_\mu,y) = \phi(x_\mu) \psi(y)$, giving a system of differential equations: 
\begin{equation}
\label{eq:freescalar2}
\left \{
\begin{array}{l}
\Box_4 \phi(x_\mu) = -\lambda^2 \phi(x_\mu) \\
 \\
\frac{d^2}{dy^2} \psi(y)  = -\lambda^2 \psi(y)
\end{array}
\right .
\end{equation}
The solution of the equation for the field $\phi(x_\mu)$ is just a plane wave, $\phi \sim e^{\pm i p \cdot x}$, with $p_\mu$ the 4-momentum and $p_\mu p^\mu = \lambda^2$.
A solution to the equation for $\psi(y)$ is given by:
\begin{equation}
\psi(y) = A e^{i k y} + B e^{- i k y}
\end{equation}
with $k^2 = \lambda^2$ the 5th-component of the 5-momentum. The 5-dimensional dispersion relation is trivially satisfied: 
$0 = \lambda^2 - k^2 = \omega^2 - |\vec{p}|^2 -k^2$, with $\omega$ the energy of $\Phi$-eigenstate with spatial momentum $(\vec{p},k)$.
Since $\psi(y)$ must be a periodic function in $y$ with period $2 \pi R$, $k$ is quantized in units of $R^{-1}$, $k = n/R$, with $n$ integer.
The energy levels of the $\Phi$ field eigenstates are, therefore, 
\begin{equation}
\omega_n = \sqrt{|\vec{p}|^2 + (n/R)^2}
\end{equation}
with $\vec{p}$ the 3-dimensional spatial component of the momentum.
If we consider $k$ as a mass term, we see that the field $\Phi$ can be decomposed into an infinite tower of massive 4-dimensional scalar fields of mass
$m_n = (n/R)$. These modes are known as Kaluza-Klein modes \cite{Kaluza:1921tu,Klein:1926tv}.

Any field configuration in this space-time can be decomposed over a set of even and odd eigenfunctions. After normalization:
\begin{equation}
\label{eq:nobraneseigenfunctions}
\left \{ 
\begin{array}{lll}
\psi^o_n (y) &=& c^o_n \sin \left ( \frac{n y}{R} \right ) \qquad {\rm for} \, n = 1, 2, \dots \\
\\
\psi^e_n (y) &=& c^e_n \cos \left ( \frac{n y}{R} \right ) \qquad {\rm for} \, n = 0, 1, \dots
\end{array}
\right .
\end{equation}
with normalization coefficients:
\begin{equation}
\label{eq:normalizationcoefficientsnobrane}
\left \{
\begin{array}{lll}
c^e_n &=& c^o_n = \frac{1}{\sqrt{\pi R}}  \qquad {\rm for} \, n = 1, 2, \dots \\
\\
c^e_0 &=& \frac{1}{\sqrt{2 \pi R}} 
\end{array}
\right .
\end{equation}

Notice that the even and odd eigenfunctions form two independent and mutually orthogonal basis for even and odd functions in ${\cal M}_4 \times {\cal S}_1$. This property
will be retained when introducing the brane, as I will show in Sect.~\ref{sec:eigenmodes}.

\subsection{Massless scalar field in ${\cal M}_4 \times {\cal S}_1$ coupled to a brane}
\label{sec:withbrane}

In Tab.~\ref{tab:operators} I review the classical dimensions of the lowest-lying local operators made out of the field $\Phi$, its derivatives and the defect
that can be added to eq.~(\ref{eq:scalaraction}) for the case of $n =1, 2$ extra-dimensions. The first column gives just the dimension of a scalar field $\Phi$ in $D=5$ or $D=6$
dimensions. Any operator with only one power of the field added to  eq.~(\ref{eq:scalaraction}) will just redefine the source of the field in the path integral formulation of
the corresponding quantum field theory and I will not consider it here. The next operator to be considered is $\Phi^2$. A mass term for the field $\Phi$ can certainly be added
to  eq.~(\ref{eq:scalaraction}) without any significant change with respect to the results of the previous section. This is not the case for the operator $\delta^{(n) (\vec y)} \, \Phi^2$. 
This operator has classical dimension 4 (6) in $n=1$ ($n=2$) extra-dimensions. If added to the lagrangian, its coefficient $q$ would have dimension 1 (0) in $n=1$ ($n=2$) 
extra-dimensions. Adding this operator to  eq.~(\ref{eq:scalaraction}) is what I call {\em minimal} coupling between the bulk field $\Phi$ and the brane, 
as it represents the only {\em relevant} ({\em marginal}) interaction that can be added in 5 (6) dimensions.
Consider, after, the operators in the third column of Tab.~\ref{tab:operators}: the term $\Phi^3$ is a self-interaction of the field and it can be safely
added to  eq.~(\ref{eq:scalaraction}). It is renormalizable both in $n=1$ and $n=2$ extra-dimensions, and it will introduce a non-trivial dynamics in the field theory. In order
to compute the KK-spectrum of the theory, however, it must be turned off to study the asymptotic free theory described by  eq.~(\ref{eq:scalaraction}).
The brane-localized term $\delta^{(n)} (\vec y) \, \Phi^3$, on the other hand, has dimension 11/2 (8) in $n=1$ ($n=2$) extra-dimensions. From an effective field theory point
of view, thus, this operator is irrelevant. Its brane-to-bulk coupling $q^\prime$ has dimension $-1/2$ ($-2$)  in $n=1$ ($n=2$) extra-dimensions. The operators of the fourth
and fifth columns of Tab.~\ref{tab:operators} are all irrelevant operators (with the notable exception of the kinetic term $\partial \Phi \partial \Phi$, of course, that is a marginal
operator). They will thus introduce suppressed non-renormalizable corrections to the physical phenomena that occur at a scale below the fundamental scale of the new physics. 
This is, in particular, the case of BLKT (fifth column of Tab.~\ref{tab:operators}), whose brane-to-bulk coupling has dimension $-1$ ($-2$) in $n=1$ ($n=2$) extra-dimensions. 
From an effective field theory point of view,  therefore, the inclusion of BLKT  and of no terms like those in the second, third or fourth columns of Tab.~\ref{tab:operators} 
is not justified\footnote{An extense literature about the inclusion of these terms into brane-world models exists, though. Particularly clear is, to my advice, the treatment
given in Ref.~\cite{delAguila:2006atw}.}.

\begin{table}
\begin{tabular}{|c|c|c|c|c|c|}
\hline
& & & & & \\
 & $\Phi $ & $\Phi^2 $ & $\Phi^3$ & $\Phi^4$ & $\partial_M \Phi \partial^M \Phi $ \\
& & & & & \\
 \hline
 $n=1$ & 3/2 & 3 & 9/2 & 6 & 5 \\
 \hline
 $n=2$ & 2 & 4 & 6 & 8 & 6 \\
\hline
& & & & & \\
& $\delta^{(n)} (\vec y) \, \Phi $ & $\delta^{(n)} (\vec y) \,  \Phi^2 $ & $\delta^{(n)} (\vec y) \, \Phi^3$ & $\delta^{(n)} (\vec y) \, \Phi^4$ & $\delta^{(n)} (\vec y) \, \partial_M \Phi \partial^M \Phi $ \\
& & & & & \\
\hline
 $n=1$ & 5/2 & 4 & 11/2 & 7 & 6 \\
 \hline
 $n=2$ & 4 & 6 & 8 & 10 & 8 \\
\hline
\end{tabular}
\caption{\it 
Classical dimensions for operators made out of powers of the $(4+n)$-dimensional scalar field $\Phi (x_\mu, y)$, its derivatives and 
the defect localized in the extra-dimension (represented by a $\delta^{(n)}(\vec y)$-function) for $n = 1$ and $n= 2$. 
Notice that $\delta^{(n)} (\vec y)$ is a dimension $n$ operator,  as obvious since $\int d^n y \, \delta^{(n)} (\vec y) = 1$ by definition.
}
\label{tab:operators}
\end{table}

On the other hand, after this short dimensional analysis, I do feel justified to consider the following action:
\begin{equation}
\label{eq:scalarplusbraneaction}
S = \int d^4x \int_{-\pi R}^{\pi R} dy \left \{ \frac{1}{2} \partial^M \Phi(x_\mu,y) \partial_M \Phi(x_\mu,y)  - \frac{q}{2} \, \delta (y) \Phi^2(x_\mu,y) \right \} 
\end{equation}
to study the impact of a {\em minimal} coupling between bulk fields and a {\em solitonic brane} \cite{Csaki:2004ay} onto the Kaluza-Klein spectrum of the theory. 
Such a  coupling of fields to a defect is indeed borrowed by the literature regarding the computation of Casimir energy in $D$-dimensions considering "realistic" conductive plates as opposed to the case where the plates are ideal ones (represented mathematically by Dirichlet boundary conditions at the points where the plates are located). 
A potential of this kind was first introduced in Ref.~\cite{Bordag:1992cm} (see also Refs.~\cite{Bordag:1998vs,Scandurra:1998xa}). Within this approach, 
the "real" conductive plates are represented by a background field $\sigma (x)$ to which the dynamical fields under study are coupled (see, {\em e.g.}, Refs.~\cite{Bordag:1992cm} 
and \cite{Graham:2003ib,Milton:2004vy}). This procedure is needed to control the divergences that usually arise in Casimir energy computation, as soon as the simplest cases 
are abandoned. Computing the Casimir energy in presence of such a background field gives a result that is finite after subtraction of a global vacuum energy divergence 
 that renormalizes the cosmological constant of the theory in the absence of boundaries. In the limit $\sigma(x) \to \delta (x)$ (called the {\em sharp limit} 
 in Ref.~\cite{Graham:2003ib}), while simultaneously taking the coupling of the field with the background to infinity ({\em strong limit}), Dirichlet boundary 
 conditions are recovered. The same approach has been subsequently applied by the MIT group in a sequence of papers, Refs.~\cite{Graham:2002xq,Graham:2002fw,Graham:2003ib,Weigel:2003tp}. More recent results can be found, for example, in Refs.~\cite{Milton:2004vy,Bordag:2004rx,Khusnutdinov:2005fi,Barone:2008hq,Guilarte:2010xn,Parashar:2012it,Bartolo:2014eoa}. 

As stressed above, the brane-to-bulk coupling $q$ in eq.~(\ref{eq:scalarplusbraneaction}) has dimension 1. The only physical scale in the model being the brane tension $f$, it is convenient to normalize the brane-to-bulk coupling as $q = c \, f$, with $c$ a dimensionless coefficient whose value is to be fixed. 
Notice that $c$ can assume any real value: for positive values, we can speak of a {\em wall} located in the 5th-dimension; for negative values,  
we have a {\em well} (or, more precisely, a {\em crevasse}).  
Higher-order operators in $\Phi$ or its derivatives are suppressed by powers of $f$ and irrelevant in the limit of large $f$. 

The idea of introducing branes as peculiar energy profiles in the bulk is not new to the realm of phenomenology, neither: in their seminal paper \cite{Randall:1999ee}, L. Randall and R. Sundrum claim that they "{\em take into account the effect of the branes on the bulk gravitational metric}" and that  they "{\em work out the consequences of the localized energy density peculiar to the brane set-up}".  As they have stated quite clearly, their results require "{\em nothing beyond the existence of 3-branes in 5 dimensions}". 
Another similar approach was pursued in Refs.~\cite{Fujimoto:2012wv,Fujimoto:2013ki,Fujimoto:2014fka,Cai:2015jla} 
by introducing boundary conditions (called {\em point interactions}) at particular points of a segment with Dirichlet boundary conditions on the edges 
(this model remember, in a sense, the UED model but with the possibility to explain the fermion mass hierarchies and flavour structure using properly tuned boundaries).

The equation of motion for the 5-dimensional field $\Phi$ is: 
\begin{equation}
\label{eq:branescalar}
\Box_5 \Phi(x_\mu,y) + q \, \delta(y) \Phi(x_\mu,y) =  0 \, .
\end{equation}
Since the $\delta$-function only affects the $y$-dependence of $\Phi(x_\mu,y)$, we can still decompose a solution to eq.~(\ref{eq:branescalar}) into
a $4$- and an extra-dimensional component, getting:
\begin{equation}
\label{eq:branescalar2}
\left \{
\begin{array}{l}
\Box_4 \phi(x_\mu) = -\lambda^2 \phi(x_\mu) \, , \\
\\
\frac{d^2}{dy^2} \psi(y) - q \, \delta(y) \psi(y) = -\lambda^2 \psi(y) \, .
\end{array}
\right .
\end{equation}
As before, the $\phi(x_\mu)$ field is just represented by plane waves, $\phi \sim e^{\pm i p \cdot x}$, with $p_\mu$ the 4-momentum and $p_\mu p^\mu = \lambda^2$,
with $\psi(y)$ the solution of the second equation in the system (\ref{eq:branescalar2}).

\section{Eigenmodes and energy levels}
\label{sec:eigenmodes}

A stationary solution for $\psi(y)$ can be put in the form: 
\begin{equation}
\label{eq:stationarysol}
\left \{
\begin{array}{ll}
\psi_1 (y) = A_1 e^{i k y} + B_1 e^{- i k y} & \qquad {\rm for } \;  y \in \, [-L/2,0[  \, , \\
\\
\psi_2 (y) = A_2 e^{i k y} + B_2 e^{- i k y} & \qquad {\rm for } \;  y \in \,  ]0,L/2] \, ,
\end{array}
\right .
\end{equation}
where $k^2 = \lambda^2$ is the 5th-component of the 5-momentum and $L = 2 \pi R$. 
The matching conditions at the position of the $\delta$-function are: 
\begin{equation}
\label{eq:matchingconditions}
\left \{
\begin{array}{l}
\psi_1(0) = \psi_2 (0) \, , \\
\\
\lim_{\epsilon \to 0} \int_{-\epsilon}^\epsilon dy \left \{ \frac{d^2}{dy^2} + k^2  - q \, \delta(y) \right \}  \psi(y) = 0 \, ,
\end{array}
\right .
\end{equation}
from which: 
\begin{equation}
\left \{
\begin{array}{l}
A_1 + B_1 = A_2 + B_2 \, , \\
\\
i k \left [ \left ( A_2 - B_2 \right ) - \left ( A_1 - B_1 \right ) \right ] = q \left ( A_1 + B_1 \right ) \, .
\end{array}
\right .
\end{equation}

The periodic boundary conditions imply that\footnote{It could be interesting to compute the impact of applying twisted boundary 
conditions at this point, \cite{Fujimoto:2013ki,Asorey:2013wca}.}: 
\begin{equation}
\left \{ 
\begin{array}{l}
\psi_1(-L/2) = \psi_2 (L/2) \longrightarrow A_1 e^{- i k L/2} + B_1 e^{i k L/2} = A_2 e^{ i k L/2} + B_2 e^{- i k L/2}  \, , \\
\\
\psi^\prime_1(-L/2) = \psi^\prime_2 (L/2) \longrightarrow A_1 e^{- i k L/2} - B_1 e^{i k L/2} = A_2 e^{ i k L/2} - B_2 e^{- i k L/2} \, .
\end{array}
\right .
\end{equation}
For a solution to exist, $k$ must be such that $\mathcal M \times \vec{A} = \vec{0}$, where $\vec{A} = (A_1, B_1, A_2, B_2)$ and
\begin{equation}
\mathcal M = 
\left ( 
\begin{array}{cccc}
1 & 1 & -1 & -1 \\
-(q + i k) & -(q - i k) & i k & - i k \\
e^{- i k L/2} & e^{+ i k L/2} & - e^{ +i k L/2} & - e^{- i k L/2} \\
e^{- i k L/2} & - e^{+ i k L/2} & - e^{ +i k L/2} & e^{- i k L/2}
\end{array}
\right ) \, .
\end{equation}
The condition for a non-trivial solution is:
\begin{equation}
\det \mathcal M = 0 \longrightarrow - 8 i k + 8 i k \cos L k + 4 i q \sin L k = 0 \, .
\end{equation}
The transcendental equation associated to the problem at hand is, then: 
\begin{equation}
\label{eq:transcendental1}
k = k \cos Lk + \frac{q}{2} \sin L k \, .
\end{equation}
Introducing the adimensional variable $\xi = L k/2 $, we can cast this equation as follows: 
\begin{equation}
\label{eq:transcendental2}
\sin \xi \left [ \xi  \sin \xi -\alpha \cos \xi \right ] = 0 \, , 
\end{equation}
with $\alpha = \pi q R/2$ (remember that $L = 2 \pi R$). This is a very important point: it can be shown that all the physical consequences derived in the rest
of the paper do not depend on the brane tension $f$ nor on the adimensional coefficient $c$ introduced above, but only on the adimensional quantity $\alpha$
that is proportional to the product of the brane tension with the compactification radius, $\alpha \propto f \, R$. This is the parameter that deserves the name
of brane-to-bulk coupling, thus. In what follows, weak and strong coupling expansion of the results are performed in terms of $\alpha$ and correspond
to the $\alpha \ll 1$ and $\alpha \gg 1$ limits, respectively.

\subsection{Odd modes}
\label{sec:oddmodes}

Eq.~(\ref{eq:transcendental2}) admits one trivial set of roots corresponding to $\xi = \pi n $ ({\em i.e.} $k_n = n/R$) for $n \geq 1$. 
The associated eigenmodes are easily found. For $\xi = \pi n$, the system to be solved reduces to: 
\begin{eqnarray}
\left \{
\begin{array}{l}
i k \left [ \left ( A_2 - B_2 \right ) - \left ( A_1 - B_1 \right ) \right ] = q \left ( A_1 + B_1 \right ) \, , \\
A_1 + B_1 = A_2 + B_2 \,  , \\
A_1 - B_1 = A_2 - B_2 \, .
\end{array}
\right .
\end{eqnarray}
From the last two conditions follows that $A_1 = A_2 \, ; B_1 = B_2$. Plugging this into the first condition gives
(for $q \neq 0$) $A_1 + B_1 = 0$: the eigenmodes vanish at $y = 0$ and are odd under $y \to - y$. After normalization:
\begin{equation}
\label{eq:oddmodes}
\psi^o_n (y) = \frac{1}{\sqrt{\pi R}} \sin \left ( \frac{n y}{R} \right ) \qquad {\rm for} \, n = 1, 2, \dots
\end{equation}
{\em i.e.} the same odd modes that are found when compactifying on a ring. This is an obvious result: since odd modes must necessarily vanish at $y = 0$, 
the discontinuity between derivatives of the eigenfunction to the left and to the right of the $\delta$-potential vanishes and the eigenmodes do not ``{\em see}'' 
the brane at all.

\subsection{Even modes}
\label{sec:evenmodes}

For $\xi \neq \pi n$ the transcendental equation to be solved is:
\begin{equation}
\label{eq:trascenonebrane}
\xi \tan \xi = \alpha \, ,
\end{equation}
whose roots have been found long ago by Burniston and Siewert \cite{PSP:2073132,Burniston:1973}:
\begin{equation}
\label{eq:xitanxialphasolutions}
\left \{ 
\begin{array}{l}
\xi_0 = \pm \sqrt{\frac{\pi \alpha}{2}} \, {\rm exp}\left \{- \frac{1}{\pi} \int_0^1 dt \frac{1}{t} \left [ \arg \Lambda_0^+ (t,\alpha) + \frac{\pi}{2} \, {\rm sign} (\alpha) \right ] \right \} \, ,\\
\\
\xi_n = \pm \frac{\pi}{2}\sqrt{4 n^2 -1 } \, {\rm exp} \left \{- \frac{1}{\pi} \int_0^1 dt \frac{1}{t} \arg \Omega_n^+ (t,\alpha) \right \} \, ,
\end{array}
\right .
\end{equation}
where
\begin{equation}
\Lambda_0^+ (t,\alpha) = \lambda (t,\alpha) - i \frac{ \pi t}{2 \alpha}  \, ,
\end{equation}
and 
\begin{equation}
\Omega_n^+ (t,\alpha) = \left [ \Lambda_0^+(t,\alpha) \right ]^2 + \frac{\pi^2 n^2}{\alpha^2} t^2 = \left \{ \lambda^2(t,\alpha) + \frac{\pi^2 d_n^2}{\alpha^2} \, t^2 \right \}
-  i \, \frac{\pi t}{\alpha}  \lambda (t,\alpha) \, .
\end{equation}
The arguments of $\Lambda_0^+ (t,\alpha)$ and $\Omega_n^+ (t,\alpha)$ are always understood to represent the principal value of the argument, {\em i.e. $\arg \Lambda_0^+ (t), \arg \Omega_n^+ (t) \in [- \pi, \pi[$}. The function $\lambda (t,\alpha)$ is:
\begin{equation}
\lambda(t,\alpha) = t \left [ t - \frac{1}{2 \alpha} \ln \frac{1-t}{1+t} \right ] \, ,
\end{equation}
and the coefficient $d_n$ is $d_n^2 = n^2-1/4$. Some details on the Burniston-Siewert method to solve analytically transcendental equations 
(useful for further application in 
compact space-times\footnote{For example, the same transcendental equation can be found in Ref.~\cite{Dienes:1998sb}, whose roots represent
the eigenvalues of the neutrino mass matrix for a SM left-handed neutrino coupled to a right-handed neutrino in a $(4+n)$-dimensional space-time compactified on an orbifold.}) are given in the Appendix. Introducing the following integrals: 
\begin{equation}
\left \{
\begin{array}{l}
I_0^\pm (\alpha) = - \frac{1}{\pi} \int_0^1 dt \frac{1}{t} \left [ \arg \Lambda_0^+ (t,\alpha) + \frac{\pi}{2} \, {\rm sign} (\alpha) \right ]  \, , \\
\\
I_n (\alpha) = - \frac{1}{\pi} \int_0^1 dt \frac{1}{t} \arg \Omega_n^+ (t,\alpha) \, ,
\end{array}
\right .
\end{equation}
(where $\pm$ refers to the sign of $\alpha$) the Kaluza-Klein masses are given by the following expressions:
\begin{equation}
\label{eq:BurnistonSiewertmodes}
\left \{ 
\begin{array}{l}
k_0^\pm (\alpha, R) = \frac{1}{R}\sqrt{\frac{\alpha}{2 \pi}} \, e^{I_0^\pm (\alpha)} \, ,\\
\\
k_n (\alpha, R) = \frac{d_n}{R} \, e^{I_n (\alpha)} \, .
\end{array}
\right .
\end{equation}
As expected, the mass of the Kaluza-Klein modes, $k_n$, is proportional to $1/R$ also in presence of a brane. However, the spectrum is quite
different from the case in which the brane is absent, as it will be shown below.

The eigenfunctions associated to the Burniston and Siewert roots are even under $ y \to - y$.  For any $n$ (but for $n=0$, $\alpha$ negative): 
\begin{equation}
\left \{ 
\begin{array}{lll}
A^n_1 &=&  e^{i k_n L} A^n_2 \, , \\
B^n_1 &=& A^n_2 \, , \\
B^n_2 &=& A^n_1 \, .
\end{array}
\right .
\end{equation}
Imposing the normalisation condition, 
\begin{equation}
\int_{- \pi R}^{\pi R} dy | \psi^e_n (y) |^2 = 1 \, ,
\end{equation}
the even eigenmodes are given by:
\begin{equation}
\label{eq:evenmodes}
\psi_n^e (y) = \frac{h_n }{\sqrt{\pi R}} \cos  k_n (|y|- \pi R)
\end{equation}
where
\begin{equation}
h_n  = \left ( 1 +  \frac{\sin 2 \pi R k_n }{2 \pi R k_n }\right )^{-1/2} \qquad {\rm for} \, 
\left \{
\begin{array}{ll}
n = 0 & {\rm for } \; \alpha > 0 \\
n = 1, 2, \dots & {\rm for \, any} \; \alpha
\end{array}
\right .
\end{equation}
after rotating away an unphysical global phase $e^{i \pi R k_n}$. 

Eq.~(\ref{eq:evenmodes}) does not apply to the case of the would-be zero-mode for negative $\alpha$, $\psi^e_{0^-}(y)$. In this case, 
eq.~(\ref{eq:BurnistonSiewertmodes}) points out that $k^-_0 (\alpha)$ is purely imaginary for any (non-vanishing) value of $|\alpha|$. 
As a consequence, the even would-be zero-mode gets stuck 
within the brane\footnote{ This property was also observed in the second Randall-Sundrum paper \cite{Randall:1999vf} for the graviton zero-mode 
in presence of a brane at $y=0$.}, that behaves like a {\em crevasse}, indeed.
The wave function for $\psi_{0^-}^e$ is given by: 
\begin{equation}
\label{eq:negativezeromode}
\psi^e_{0^-} (y) = \frac{h_{0^-} }{\sqrt{\pi R}} \cosh |k_{0^ -}| (|y|- \pi R) \, ,
\end{equation}
with normalisation coefficient:
\begin{equation}
h_{0^-}  = \left ( 1 +  \frac{\sinh 2 \pi R |k_{0^-} |}{2 \pi R |k_{0^-} |}\right )^{-1/2} \, .
\end{equation}

\subsection{Properties of eigenmodes and eigenvalues}
\label{sec:properties}

Even and odd eigenmodes, given in eqs.~(\ref{eq:evenmodes}) and (\ref{eq:oddmodes}) respectively, form two independent orthonormal 
bases\footnote{This statement can be easily proved by using eq.~(\ref{eq:trascenonebrane}).}, mutually orthogonal, for even and odd functions in 
the extra-dimensional coordinate $y$:
\begin{equation}
\left \{
\begin{array}{l}
< \psi^e_n (y) | \psi^e_m (y) > = < \psi^o_n(y) | \psi^o_m (y) > = \delta_{nm} \\
< \psi^e_n (y) | \psi^o_m (y)> = 0
\end{array}
\right .
\end{equation}

\begin{figure}[htb!]
\begin{center}
	\includegraphics[width=14cm]{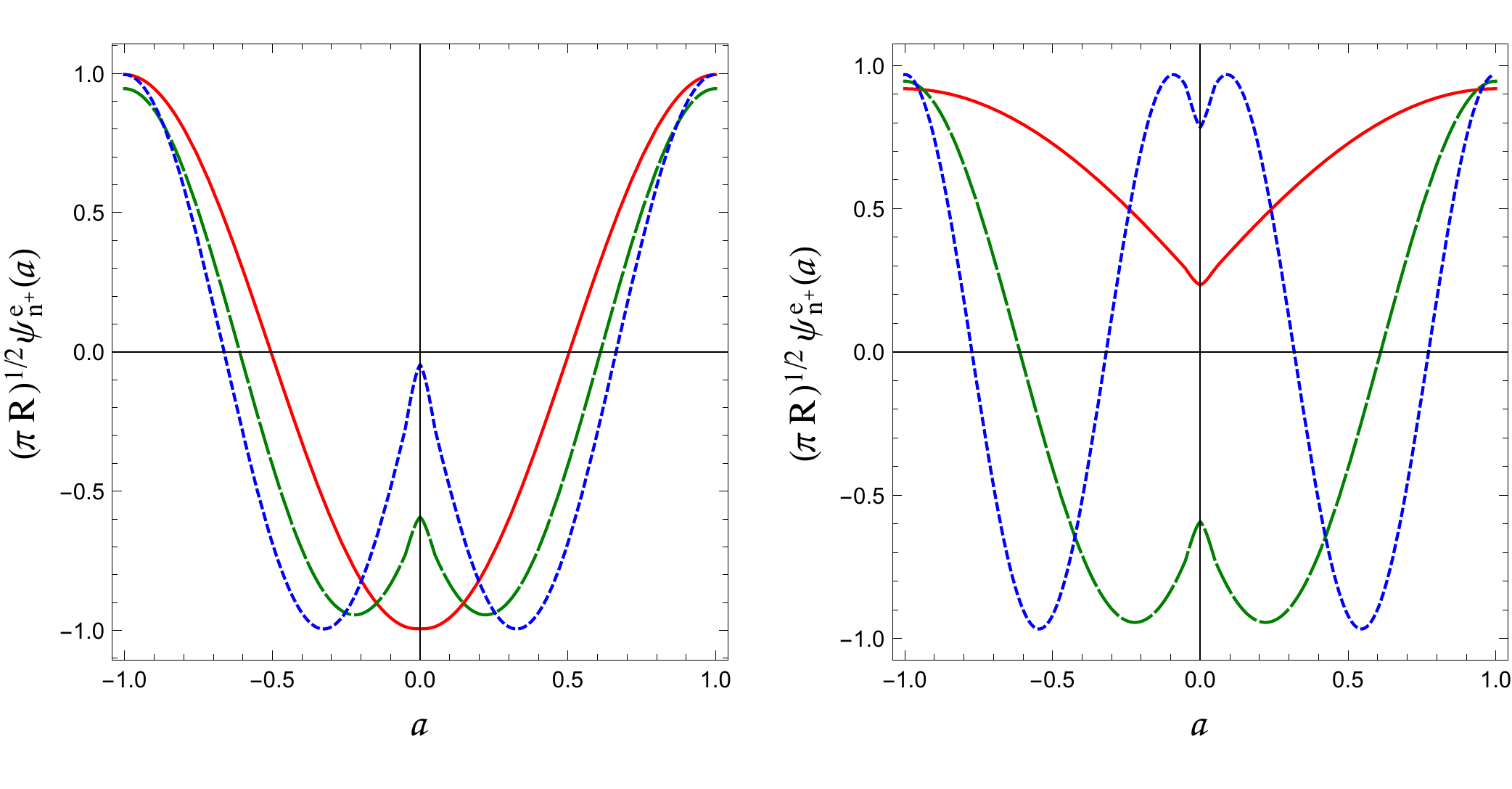}
\caption{\it Left: The dependence of $\psi_1^e$ on the brane-to-bulk coupling $|\alpha|$ for three representative values, $\alpha = + 0.1$ (solid red); 
$\alpha = + 5$ (dashed green) and $\alpha = + 100$ (dotted blue), as a function of the coordinate $a = y/ \pi R$. 
Right: the dependence of $\psi_n^e$ on $n$, for $n = 0$ (solid red), $n = 1$ (dashed green) and $n=2$ (dotted blue) for $\alpha =+5$, as a function of the coordinate $a = y/ \pi R$. 
 }
\label{fig:depwavefunction}
\end{center}
\end{figure}

The dependence of the even eigenmodes on $\alpha$ and $n$, as a function of the (normalized) extra-dimensional coordinate $a = y/(\pi R)$, is depicted in Fig.~\ref{fig:depwavefunction} (left and right panels, respectively).

\begin{figure}[htb!]
\begin{center}
	\includegraphics[width=14cm]{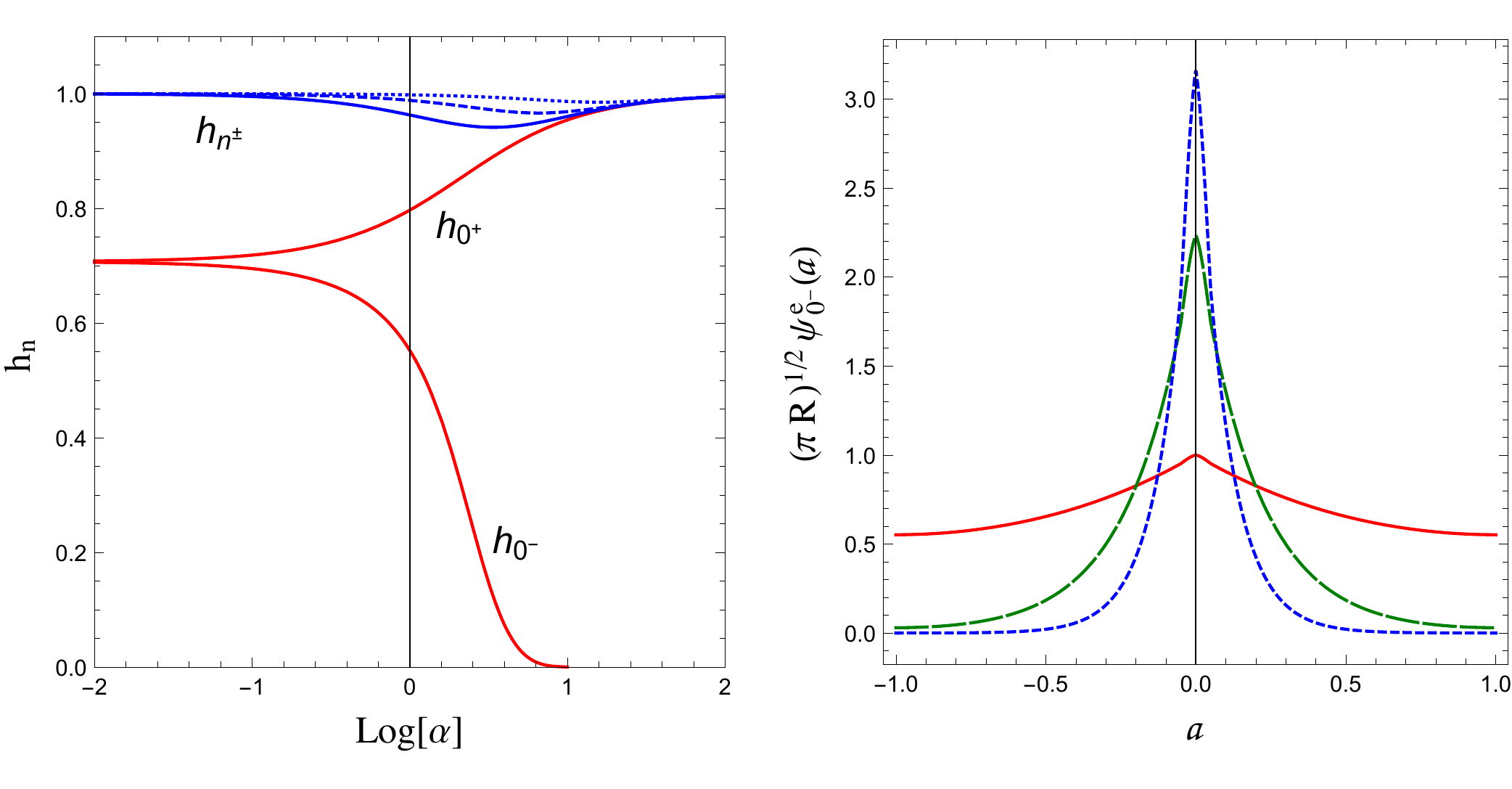}
\caption{\it Left: The dependence of $h_n$ on the brane-to-bulk coupling $|\alpha|$ for $n = 0^+, 0^-$ (red), $n = 1$ (solid,blue), $n=2$ (dashed, blue) and $n=5$ (dotted, blue). 
Right: the profile of the would-be zero-mode $\psi_{0^-}^e$ in the extra-dimension for $\alpha = - 1$ (solid, red), $- 5$ (dashed, green) and $- 10$ (dotted, blue). Notice that for $\alpha = -1$ the mode $\psi^e_{0^-}$ is not fully localized EXPAND ON THIS CONCEPT.
 }
\label{fig:depcoefficient}
\end{center}
\end{figure}

For any non-vanishing $n$, the limit of the normalization coefficients is $h_n \to 1$ both for $|\alpha| \to 0$ and 
$|\alpha| \to \infty$. The functional form of eq.~(\ref{eq:evenmodes}) is also valid for the would-be zero-mode $\psi^e_{0^+}(y)$ for positive $\alpha$, albeit with
$h_{0^+} \to 1/\sqrt{2}$ for $|\alpha| \to 0$ (as it should, see eq.~(\ref{eq:normalizationcoefficientsnobrane})).
The dependence of $h_n$ and $h_{0^+}$ from $|\alpha|$ is shown in Fig.~\ref{fig:depcoefficient} (left). 

In Fig.~\ref{fig:depcoefficient} (right) the profile of $\psi_{0^-}^e$ as a function of $a = y/ (\pi R)$ for several values of $|\alpha|$ is shown. Notice that, for increasing $|\alpha|$, 
the extension of the zero-mode into the bulk decreases exponentially fast. For relatively large values of $|\alpha|$, the mode is effectively confined into the brane. Localisation of 
a field in a region of the space-time when the field is coupled to some defect is not new: fermions coupled to a kink, for example, are known to get localised in the vicinity
of the discontinuity. This mechanism was used in Ref.~\cite{ArkaniHamed:1999dc} to build a model in which Yukawa couplings are computed as the overlap of fermion wavefunctions with a background scalar kink whose profile is smooth in the extra-dimension. Similar mechanisms to localise fermions were proposed in Refs.~\cite{Dvali:2000ha,Kaplan:2001ga}.
Gauge fields, also, have been found to localise in models with non-trivial background scalar field: see, for example, Refs.~\cite{Dvali:1996xe,Dvali:2000bz,Dvali:2000rx}.
Needless to say, a model in which all of the SM fields (scalars, fermions and gauge bosons) localise into a brane, whereas gravity does not, is precisely the target
model needed to solve the hierarchy problem, as discussed in the Introduction.

\begin{figure}[htb!]
\begin{center}
	\includegraphics[width=14cm]{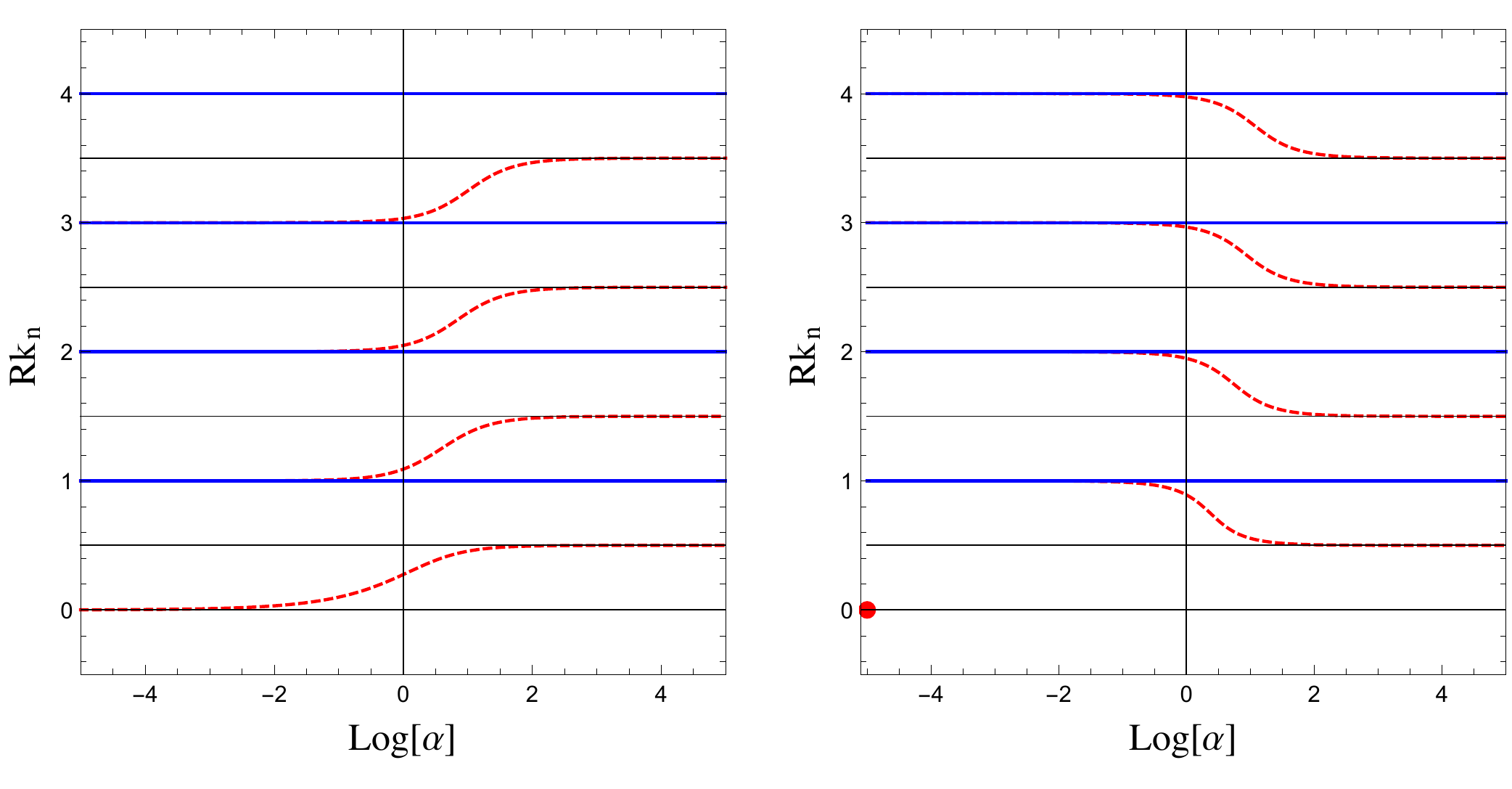}
\caption{\it The Kaluza-Klein spectrum for a scalar field in ${\cal M}_4 \times {\cal S}_1$ (with compactification radius $R$) in presence of a brane located at $y = 0$
as a function of the brane-to-bulk coupling $|\alpha|$.
Left: the case for a potential wall, $\alpha > 0$. Right: the case for a potential crevasse, $\alpha < 0$.
Solid (blue) lines represent the eigenvalues associated to odd modes; dashed (red) lines stand for the $\alpha$-dependent eigenvalues associated to even modes;
thin black lines stand for the asymptotic eigenvalues $R k_n = (n \pm 1/2)$ for even modes in the limit $\alpha \to \pm \infty$.
In the case of negative $\alpha$, the lowest-lying eigenvalue becomes imaginary (i.e., the corresponding mode is localised in the extra-dimension, but for $\alpha= 0$, 
represented by a red dot in the right panel).
 }
\label{fig:spectrumonebrane}
\end{center}
\end{figure}

The Kaluza-Klein spectrum is eventually shown in Fig.~\ref{fig:spectrumonebrane}  as a function of $|\alpha|$ for positive $\alpha$ (left panel) and negative $\alpha$ (right panel).
In both cases, in the limit $\alpha \to 0$ the standard spectrum for the Kaluza-Klein modes of a scalar field in the absence of the brane, $k_n = n/R$, is recovered. 
However, for any value of $\alpha \neq 0$ the spectrum is modified. It is easy to see that, for positive $\alpha$ (left panel), all modes are shifted at higher energies 
by the presence of the brane and, in particular, the lightest mode $\psi^e_{0^+}$ (with vanishing mass in the limit $\alpha = 0$) gets a mass $k^+_0 (\alpha)$.
On the other hand, for negative $\alpha$ (right panel), all modes with $n \geq 1$ are shifted to lower energies, whereas the would-be zero-mode $\psi^e_{0^-}$ 
(represented by a red dot) gets secluded onto the brane for any non-vanishing $\alpha$.

\subsection{Comparison with Brane-Localized Kinetic Terms}
\label{sec:BLKT}

Consider the action \cite{Dvali:2001gm}
\begin{equation}
\label{eq:BLKTaction}
S = \int d^4x \int_{-\pi R}^{\pi R} dy \left \{ \frac{1}{2} \partial^M \Phi(x_\mu,y) \partial_M \Phi(x_\mu,y)   +  r_c \, \delta (y) \, \partial^\mu \Phi(x_\mu,y) \partial_\mu \Phi(x_\mu,y) \right \}  \, ,
\end{equation}
where the second term is a correction to the kinetic term of the bulk field $\Phi (x_\mu, y)$ localized at the position of the brane at $y=0$ (a {\em brane-localized kinetic term}). 
Its origin may be ascribed to loops of fields localized into the brane. The parameter $r_c$ is the length scale beyond which 4-dimensional gravity is modified (as the localized fields
also induce a local curvature tensor $R$ that may modify gravity in the bulk \cite{Dvali:2001gm}). It is not this the place to comment on the motivation of adding this (irrelevant, as stressed above) operator to eq.~(\ref{eq:scalaraction}). What is interesting, however, is that the Burniston-Siewert method outlined in the Appendix allows us to immediately 
derive an exact expression for the Kaluza-Klein spectrum of the model. The equation of motion for the 5-dimensional field $\Phi$ is: 
\begin{equation}
\label{eq:branescalarBLKT}
\Box_5 \Phi(x_\mu,y) + r_c \, \delta(y) \Box_4 \Phi(x_\mu,y) =  0
\end{equation}
and it can be decomposed as follows: 
\begin{equation}
\label{eq:branescalarBLKT2}
\left \{
\begin{array}{l}
\Box_4 \phi(x_\mu) = -\lambda^2 \phi(x_\mu) \, , \\
\\
\frac{d^2}{dy^2} \psi(y) + \lambda^2 \left [ 1 + r_c \, \delta(y) \right ] \psi(y) = 0.
\end{array}
\right .
\end{equation}
As before, the $\phi(x_\mu)$ field is just represented by plane waves, $\phi \sim e^{\pm i p \cdot x}$, with $p_\mu$ the 4-momentum and $p_\mu p^\mu = \lambda^2$,
with $\psi(y)$ the solution of the second equation in the system (\ref{eq:branescalarBLKT2}).

Also in this case, a stationary solution for $\psi(y)$ to the left and to the right of the $\delta$-function can be expressed as in eq.~(\ref{eq:stationarysol}), with
$k^2 = \lambda^2$ the 5th-component of the 5-momentum and $L = 2 \pi R$. After imposing periodic boundary conditions at $y = - L/2$ and $y = L/2$ and 
matching conditions at $y= 0$, a non-trivial solution in $k$ is found solving $\mathcal M_{\rm BLKT} \times \vec{A} = \vec{0}$, where $\vec{A} = (A_1, B_1, A_2, B_2)$ and
\begin{equation}
\mathcal M_{\rm BLKT} = 
\left ( 
\begin{array}{cccc}
1 & 1 & -1 & -1 \\
k^2 r_c - i k & k^2 r_c + i k & i k & - i k \\
e^{- i k L/2} & e^{+ i k L/2} & - e^{ +i k L/2} & - e^{- i k L/2} \\
e^{- i k L/2} & - e^{+ i k L/2} & - e^{ +i k L/2} & e^{- i k L/2}
\end{array}
\right ) \, .
\end{equation}
In order to have a non-trivial solution, 
\begin{equation}
\det \mathcal M_{\rm BLKT} = 0 \longrightarrow - 8 i k + 8 i k \cos L k - 4 i r_c k^2  \sin L k = 0 \, .
\end{equation}
The relevant transcendental equation to be solved is, then: 
\begin{equation}
\label{eq:transcendentalBLKT1}
k = k \cos Lk - \frac{r_c k^2}{2} \sin L k \, .
\end{equation}
Introducing the adimensional variable $\xi = L k/2 $, we can cast this equation as follows: 
\begin{equation}
\label{eq:transcendentalBLKT2}
(\xi \sin \xi )\left [ \sin \xi - \alpha \xi \cos \xi \right ] = 0 \, , 
\end{equation}
with\footnote{ I have defined $\alpha$ in this way to match with the notation given in the Appendix.} 
$\alpha = - r_c/L = - r_c/(2 \pi R)$. Also in this case, the spectrum can be split into two sectors: 
\begin{equation}
\label{eq:transcendentalBLKT3}
\left \{
\begin{array}{lll}
\xi \, \sin \xi = 0 & \longrightarrow & {\rm odd \; modes}: \; \xi = n \pi \, ,
\\
\tan \xi = \alpha \xi & \longrightarrow & {\rm even \; modes} \, .
\end{array}
\right .
\end{equation}
The first set of solutions are nothing more than the sine-functions corresponding to modes that are odd with respect to the position of the brane. As usual, as they do not
see the brane at $y=0$ (having a node there), they are not affected at all by the presence of the brane.  Notice that this set of modes, usually neglected in the literature
on BLKT, cannot interact with brane fields at tree level. They can, and in general will, interact with brane modes beyond the tree level if bulk fields are self-interacting
(as it is, for example, the case for gravitons). 

The second set of solutions, corresponding to modes even with respect to the position of the brane, was studied numerically in Ref.~\cite{Dvali:2001gm}. 
However, we can immediately recognize that the second equation in (\ref{eq:transcendentalBLKT3}) is one of the transcendental equation solved by Burniston 
and Siewert in Ref.~\cite{PSP:2073132}. In the Appendix I explain how to derive an exact solution for this equation. I summarize here the results: 
first of all, notice that $k=0$ is a trivial solution of eq.~(\ref{eq:transcendentalBLKT2}) for any value of $\alpha$ and that, therefore, a constant zero-mode is always
present in the spectrum. On the other hand,  for $\alpha = - r_c/L < 0$ (a straightforward consequence of assuming that both $L$ and $r_c$ are length scales and, thus, positive), 
only solutions with $n \neq 0$ exist: 
\begin{equation}
\xi_n = \pm \pi n \exp \left [ \frac{1}{\pi} \int_0^1 dt \frac{1}{t} \arg \Omega_n^+ (t) \right ] \qquad n = 1, 2, \dots,\; \alpha \in ] - \infty, + \infty [ \, ,
\end{equation}
where:
\begin{equation}
\Lambda_0^+ (t,\alpha) = \lambda (t,\alpha) + i \frac{ \pi \alpha t}{2}  \, ,
\end{equation}
and 
\begin{equation}
\Omega_n^+ (t,\alpha) = \left [ \Lambda_0^+(t,\alpha) \right ]^2 + \pi^2 n^2 \alpha^2 t^2 \, .
\end{equation}
The function $\lambda (t,\alpha)$ is, in this case:
\begin{equation}
\lambda(t,\alpha) = 1 + \frac{\alpha t}{2} \ln \frac{1-t}{1+t} \, .
\end{equation}

Notice, however, that the assumption that the coefficient $r_c$ in eq.~(\ref{eq:BLKTaction}) be strictly positive is not a consequence of a thorough
computation of the local quantum corrections to the action for a bulk field due to brane fields (that is, of course, model dependent). If we allow for $r_c$ to be
negative (and, therefore, $\alpha > 0$), then an additional even mode is found: 
\begin{equation}
\xi_0 = \pm \frac{1}{\alpha} (\alpha - 1)^{1/2} \exp \left [ \frac{1}{\pi} \int_0^1 dt \frac{1}{t} \arg \Lambda_0^+ (t) \right ] \qquad (\alpha > 0) \, .
\end{equation}
For $\alpha >1$, this mode propagates into the bulk together with the rest of the even modes $\xi_n$. On the other hand, for $\alpha \in ] 0, 1 [$, the mode gets trapped
into the brane. We see that localization of an even mode only occurs\footnote{This was first noticed in Ref.~\cite{delAguila:2003bh}.} in presence of BLKT if the brane-to-bulk coupling $\alpha$ is positive ({\em i.e.} $r_c < 0$) and 
$\alpha \in ] 0, 1[$ ($|r_c| \in ] 0, L [ $). Eventually, for the special case $\alpha = 1$, this mode overlaps the standard constant zero-mode. 

Normalization of the even modes is easily carried on along the lines of Sect.~\ref{sec:evenmodes}, finding the same results: 
\begin{equation}
\label{eq:evenmodesBLKT}
\psi_n^e (y) = \frac{h_n }{\sqrt{\pi R}} \cos  k_n (|y|- \pi R) \, ,
\end{equation}
where
\begin{equation}
\label{eq:hnBLKT}
h_n  = \left ( 1 +  \frac{\sin 2 \pi R k_n }{2 \pi R k_n }\right )^{-1/2} \qquad {\rm for} \, n = 1, 2, \dots \, ,
\end{equation}
after rotating away an unphysical global phase $e^{i \pi R k_n}$. In order to compare eq.~(\ref{eq:hnBLKT}) with the results of Ref.~\cite{Dvali:2001gm}, it suffices
to use the transcendental equation $\tan \xi = \alpha \xi$ to get: 
\begin{equation}
\label{eq:hnBLKT2}
h_n  = \left ( 1 +  \frac{\alpha}{1 + \alpha^2 \xi_n^2 }\right )^{-1/2} = \left ( 1 -  \frac{r_c/L}{1 + r_c^2 k_n^2 / 4 }\right )^{-1/2} \qquad {\rm for} \, n = 1, 2, \dots
\end{equation}

Notice that in the case of BLKT, whereas odd modes form an orthonormal basis for odd functions of $y$ and are trivially shown to be orthogonal to even modes, 
the even modes are not orthornomal between themselves. 
\begin{equation}
\left \{
\begin{array}{l}
< \psi^o_n(y) | \psi^o_m (y) > = \delta_{nm} \\
< \psi^e_n (y) | \psi^e_m (y) > \neq \delta_{nm} \\
 < \psi^e_n (y) | \psi^o_m (y)> = 0
\end{array}
\right .
\end{equation}
This is an obvious consequence of the fact that the differential operator in eq.~(\ref{eq:branescalarBLKT}) is not a Sturm-Liouville differential operator. However, 
it can be easily shown making use of the transcendental equation satisfied by the eigenvalues $\xi_n$ that first derivatives of the even eigenfunctions are indeed orthogonal
between themselves. 

\section{Weak coupling limit: $\alpha \ll 1$}
\label{sec:weakcoupling}

The weak coupling expansion represents the physical situation of a {\em semi-transparent brane},  whose presence in the extra-dimension is 
feeble and modifies the Kaluza-Klein spectrum only slightly\footnote{
Notice that this situation is the one that is less interesting from the phenomenological point of view. Since $\alpha = (c \pi/2) \, f \, R$, to have $\alpha \ll 1$ 
it must be $f \, R \ll 2/(\pi c)$. For $c = {\cal O}(1)$, this translates into $f \ll 1/R$. Since the typical size of the brane fluctuations in the bulk is ${\cal O}(f^{-1})$ 
\cite{DeRujula:2000he}, in the weak coupling limit one expects that modes confined to the brane may wrap many times around the compact dimension and effectively {\em see} it by just sitting where they are (it is the {\em soft brane} itself that bring them to take a walk into the bulk).
Kaluza-Klein modes of the SM particles with masses $m  = {\cal O}(f) \ll R^{-1}$ arise in this picture, in contradiction with experiments.}.

Consider first the modes with $n \geq 1$, $k_n (\alpha ) =  (d_n/R) \exp \left [ I_n (\alpha) \right ]$. 
Expanding formally in Taylor series $k_n(\alpha)$ for $\alpha \ll 1$: 
\begin{equation}
k_n (\alpha) \simeq \left ( \frac{d_n}{R}  \right ) \left . e^{I_n (\alpha)} \right |_{\alpha = 0} \left \{ 1 + \left ( \left . \frac{d I_n (\alpha)}{d\alpha} \right |_{\alpha = 0} \right ) \alpha + \dots
\right \} \, .
\end{equation}
It is possible to verify numerically that
\begin{equation}
\lim_{\alpha \to 0} \exp \left [ I_n(\alpha) \right ] = \frac{n}{d_n} \, ,
\end{equation}
a result obvious from both panels of Fig.~\ref{fig:spectrumonebrane}, but not at all obvious analytically. 
Using the explicit expression for the integral $I_n(\alpha)$,
\begin{equation}
\label{eq:intInalpha}
I_n (\alpha) =  \frac{1}{\pi} \int_0^1 dt \, \frac{1}{t} \arctan \left \{ \frac{\left ( \frac{\alpha \lambda}{\pi t} \right ) }{\left ( \frac{\alpha \lambda}{\pi t} \right )^2+d_n^2}
\right  \} \, ,
\end{equation}
the leading term in the expansion can be computed by taking the derivative under the integral: 
\begin{equation}
 \left . \frac{d I_n (\alpha)}{d\alpha} \right |_{\alpha = 0}  = 
 \frac{1}{\pi} \int_0^1 dt \, \frac{1}{t} 
 \left ( \left . \frac{d}{d\alpha} \arctan \left \{ \frac{\left ( \frac{\alpha \lambda}{\pi t} \right ) }{\left ( \frac{\alpha \lambda}{\pi t} \right )^2+d_n^2} \right \} \right |_{\alpha = 0} \right )
= \int_0^1 dt \, D_{n1}(t) \, ,
\end{equation}
where
\begin{equation}
D_{n1}(t) = 4 \frac{\pi^2(4 n^2-1)-\ln^2 \left( \frac{1-t}{1+t}\right ) }{\pi^4 (4 n^2-1)^2+2 \pi^2 (4 n^2+1)\ln^2 \left( \frac{1-t}{1+t}\right )+\ln^4 \left( \frac{1-t}{1+t}\right ) } \, .
\end{equation}
To get a closed-form expression, I first change variables: 
\begin{equation}
t \to (1-y)/(1+y) \, ; \qquad dt = - 2/(1+y)^2 dy \, ,
\end{equation}
such that
\begin{equation}
 \left . \frac{d I_n (\alpha)}{d\alpha} \right |_{\alpha = 0}  = 2 \int_0^1 dy \frac{1}{(1+y)^2} \, D_{n1}(y)
\end{equation}
and
\begin{equation}
D_{n1}(y) = 4 \frac{\pi^2(4 n^2-1)-\ln^2 y}{\pi^4 (4 n^2-1)^2+2 \pi^2 (4 n^2+1)\ln^2 y+\ln^4 y } \, .
\end{equation}
Then, with a second change of variables:
\begin{equation}
\ln y \to - x \, ; \qquad dy = - e^{-x} dx \, ,
\end{equation}
I get: 
\begin{equation}
 \left . \frac{d I_n (\alpha)}{d\alpha} \right |_{\alpha = 0}  = 2 \int_0^\infty dx \frac{e^{-x}}{\left (1+e^{-x} \right )^2} \, D_{n1}(x) 
\end{equation}
and
\begin{equation}
D_{n1}(x) = 4 \frac{\pi^2(4 n^2-1)-x^2}{\pi^4 (4 n^2-1)^2+2 \pi^2 (4 n^2+1) x^2+ x^4 } \, .
\end{equation}
Expanding this expression in Laurent series for large $n$: 
\begin{equation}
D_{n1}(x) = \frac{1}{\pi^2 n^2} + {\cal O} \left (\frac{1}{\pi^4 n^4} \right ) \, ,
\end{equation}
it comes out that, upon integration, only the first term of the series gives a non-vanishing contribution. Therefore, 
\begin{equation}
 \left . \frac{d I_n (\alpha)}{d\alpha} \right |_{\alpha = 0}  = 2 \int_0^\infty dx \frac{e^{-x}}{\left (1+e^{-x} \right )^2} \, D_{n1}(x) = \frac{1}{\pi^2 n^2} \, ,
\end{equation}
an expression valid for all $n \geq 1$ (the validity of this expression also for $n=1$ is not surprising, as the effective expansion parameter is $1/(\pi n)$ and not $1/n$ itself). 
Using the same strategy, the subleading terms in the Taylor expansion of $k_n(\alpha)$ in powers of $\alpha$ can be computed. 
Eventually, the result is: 
\begin{equation}
\label{eq:knweakcoupling}
k_n (\alpha) \simeq \frac{n}{R} \left \{ 1 + \frac{\alpha}{\pi^2 n^2} - \frac{\alpha^2}{\pi^4 n^4} + \dots  \right \}  \, ,
\end{equation}
where this expression is valid for both positive and negative $\alpha$ under the trivial change $\alpha \to - \alpha$.

A similar approach can be used to compute an approximate expression for the {\em would-be} zero-mode eigenvalue $k_0^+$
for $0 < \alpha  \ll 1$.  In this case, 
\begin{equation}
k_0^+ (\alpha ) =  \frac{1}{R} \sqrt{\frac{ \alpha}{2\pi}} \,  e^{I_0^+ (\alpha)} \, ,
\end{equation}
and the explicit expression for the $I_0^+$ integral is:
\begin{equation}
I_0^+ (\alpha) =  -\frac{1}{\pi} \int_0^1 dt \frac{1}{t} \left \{ \frac{\pi}{2} - \arctan \left ( \frac{\pi t}{2 \alpha \lambda} \right ) \right \} \, .
\end{equation}
Therefore,
\begin{equation}
k_0^+ (\alpha) \simeq \frac{1}{R}  \sqrt{\frac{\alpha}{2\pi}} \left . e^{I_0^+ (\alpha)} \right |_{\alpha = 0} 
\left \{ 1 + \left ( \left . \frac{d I_0^+ (\alpha)}{d\alpha} \right |_{\alpha = 0} \right ) \alpha + \dots
\right \} \, .
\end{equation}
Eventually, we get\footnote{Notice that eqs.~(\ref{eq:knweakcoupling}) and (\ref{eq:k0weakcoupling}) reproduce the approximate results given in 
eq.~(2.36) of Ref.~\cite{Dienes:1998sb}, setting $\alpha = (\pi m R)^2$ and $\xi = \pi \lambda R$.}:
\begin{equation}
\label{eq:k0weakcoupling}
k_0^+(\alpha) \simeq  \frac{\sqrt{\alpha}}{\pi R}  \, \left \{ 1 - \frac{1}{6} \alpha + \frac{11}{360} \alpha^2 + \dots \right \} \, .
\end{equation}

It can be shown that the even eigenmodes $\psi_n^e (y)$ reduce trivially to standard even standing waves in the limit $\alpha \to 0$, as it should be in the absence of the brane.

\section{Strong coupling limit: $\alpha \gg 1$}
\label{sec:strongcoupling}

This second case represents what could be called a {\em stiff brane}, whose presence modifies strongly the spectrum. It is this case that, actually, 
is more appealing from the phenomenological point of view: in order for $\alpha$ to be much greater than 1, the brane tension $f$ must be much larger than 
the inverse compactification radius, $f \gg R^{-1}$, for $c = {\cal O}(1)$. Brane fluctuations into the bulk are, therefore, completely negligible as $f^{-1} \ll R$ \cite{DeRujula:2000he} and the brane
itself behaves as a rigid object with no KK-excitations of SM fields stuck to it.

Consider first the $k_n$ modes with $n \geq 1$ for positive $\alpha$. Changing the integration variable in eq.~(\ref{eq:intInalpha}) to $x = \alpha t/\pi$ and integrating by parts:
\begin{equation}
I_n(\alpha) = - \frac{1}{\pi} \int_0^{\alpha/\pi} dx \frac{1}{x} f(x) = - \left . \frac{1}{\pi}  \ln x f(x) \right |_{0}^{\alpha/\pi} + \frac{1}{\pi} \int_0^{\alpha/\pi} dx \ln x \frac{d f(x)}{dx} \, ,
\end{equation}
where
\begin{equation}
f(x) = \left . \arctan \left \{ - \left ( \frac{\alpha \lambda}{\pi t}\right ) \left [ \left ( \frac{\alpha \lambda}{\pi t}\right )^2 + d_n^2 \right ]^{-1}  \right \} \right |_{t = (\pi /\alpha) x} \, .
\end{equation}

Expanding in Laurent series for  $\alpha \to \infty$:
\begin{equation}
I_n (\alpha) = A_n (\alpha) + J_{n1}(\alpha) +  J_{n2}(\alpha) + {\cal O}(1/\alpha^3) \, ,
\end{equation}
where
\begin{equation}
\left \{
\begin{array}{lll}
A_n (\alpha) &=& -\left . \frac{1}{\pi} \ln x f(x) \right |_{0}^{\alpha/\pi}  \, , \\
\\
J_{n1} (\alpha) &=& \frac{1}{\pi}  \int_0^{\alpha/\pi} dx \ln x \frac{(x^2-d_n^2)}{[(x^2+d_n^2)^2 + x^2]} \, , \\
\\
J_{n2} (\alpha) &=& -\frac{1}{\pi \alpha} \int_0^{\alpha/\pi} dx \ln x \frac{(x^2 + d_n^2) (x^4 - x^2 - 6 d_n^2 x^2 + d_n^4)}{[(x^2+d_n^2)^2 + x^2]^2}  \, .
\end{array}
\right .
\end{equation}
Up to second order in $1/\alpha$ we get:
\begin{equation}
\left \{
\begin{array}{lll}
A_n (\alpha) &=& - \frac{1}{\alpha} \ln  \frac{\pi}{\alpha} + \frac{1}{\alpha^2} \ln \frac{\pi}{\alpha}+ {\cal O} \left (\frac{1}{\alpha^3} \right ) \, ,\\
\\
J_{n1}(\alpha) &=& = \frac{1}{2} \ln \left ( \frac{2 n +1}{2 n -1} \right ) -\frac{1}{\alpha} \left (1 - \ln \frac{\pi}{\alpha}\right ) + {\cal O} \left (\frac{1}{\alpha^3} \right ) \, , \\
\\
J_{n2}(\alpha) &=& \frac{1}{\alpha^2}  \left (1 - \ln \frac{\pi}{\alpha}\right ) + {\cal O} \left (\frac{1}{\alpha^3} \right ) \, .
\end{array}
\right .
\end{equation}
Therefore,  
\begin{equation}
I_n (\alpha) = \frac{1}{2} \ln \left ( \frac{2 n + 1}{2 n -1} \right ) - \frac{1}{\alpha} + \frac{1}{\alpha^2} + {\cal O} \left (\frac{1}{\alpha^3} \right )
\end{equation}
Terms up to second order in $1/\alpha$ are universal for all KK-modes with $n \geq 1$, whereas an $n$-dependence arises at ${\cal O }( 1/\alpha^3)$. 
The explicit analytic expression for ${\cal O }( 1/\alpha^3)$ terms is not particularly inspiring and it will not be presented here. 

The computation for negative $\alpha$ can be carried on along similarly, to get eventually the final result for the strong coupling expansion of $k_n$ eigenvalues:
\begin{equation}
\label{eq:knstrongcoupling}
k_n (\pm |\alpha|) \simeq \frac{1}{R} \left (n \pm \frac{1}{2} \right ) \exp \left [ \mp \frac{1}{|\alpha|} + \frac{1}{|\alpha|^2} + {\cal O} \left (\frac{1}{\alpha^3} \right ) \right ] \, .
\end{equation}

The energy of the {\em would-be} zero-mode $k_0^+$ in the strong coupling limit can be computed in a similar way: 
\begin{equation}
k^+_0 = \frac{1}{R}\sqrt{\frac{\alpha}{ 2 \pi}} \, \exp \left [ I^+_0 (\alpha) \right ] = \frac{1}{2 R} \exp \left (\frac{1}{2} \ln \frac{2 \alpha}{\pi} + I^+_0 (\alpha) \right ) \, ,
\end{equation}
where
\begin{equation}
I^+_0(\alpha) = -\frac{1}{\pi} \int_0^1 dt \frac{1}{t} \left \{ \arctan \left ( - \frac{\pi t}{2 \alpha \lambda} \right ) + \frac{\pi}{2}  \right \} \, .
\end{equation}

Changing the integration variable to $x = 2 \alpha t/\pi$ and integrating by parts:
\begin{equation}
I^+_0(\alpha) = - \frac{1}{\pi} \int_0^{2\alpha/\pi} dx \frac{1}{x} \, g^+(x) = 
- \left . \frac{1}{\pi}   \ln x g^+(x)  \right |_{0}^{2\alpha/\pi} + \frac{1}{\pi} \int_0^{2 \alpha/\pi} dx \ln x \frac{d g^+(x)}{dx} \, ,
\end{equation}
where
\begin{equation}
g^+(x) = \frac{\pi}{2} - \left . \arctan \left ( \frac{\pi t}{2 \alpha \lambda} \right ) \right |_{t = (\pi/ 2\alpha) x} \, .
\end{equation}
Expanding in powers of $1/\alpha$:
\begin{equation}
I_0^+(\alpha) = A^+_0 (\alpha) + J^+_{01}(\alpha) +  J^+_{02}(\alpha)  +  J^+_{03}(\alpha) + \dots  \, ,
\end{equation}
where (up to second order in $1/\alpha$):
\begin{equation}
\left \{
\begin{array}{lll}
A^+_0 (\alpha) &=&  -\left . \frac{1}{\pi}  \ln x g^+(x) \right |_{0}^{2\alpha/\pi} = \frac{1}{2} \ln \frac{\pi}{2\alpha}  - \frac{1}{2\alpha}  \ln \frac{\pi}{2\alpha} 
+\frac{1}{2 \alpha^2} \ln \frac{\pi}{2 \alpha}
+ {\cal O} \left (\frac{1}{\alpha^3} \right ) \, , \\
\\
J^+_{01} (\alpha) &=& \frac{1}{\pi}  \int_0^{2 \alpha/\pi} dx \ln x \frac{1}{(1+ x^2)} = - \frac{1}{2 \alpha} \left ( 1 - \ln \frac{\pi}{2\alpha}\right ) + {\cal O} \left (\frac{1}{\alpha^3} \right ) \, , \\
\\
J^+_{02} (\alpha) &=& \frac{1}{\pi \alpha} \int_0^{2 \alpha/\pi} dx \ln x \frac{1- x^2}{(1+ x^2)^2}= -\frac{1}{2\alpha} 
+ \frac{1}{2 \alpha^2} \left ( 1 - \ln \frac{\pi}{2 \alpha} \right ) + {\cal O} \left (\frac{1}{\alpha^3} \right ) \, , \\
\\
J^+_{03} (\alpha) &=& -\frac{1}{\pi \alpha^2} \int_0^{2 \alpha/\pi} dx \ln x \frac{x^2 (3- x^2)}{(1+ x^2)^3} = \frac{1}{4 \alpha^2} + {\cal O} \left (\frac{1}{\alpha^3} \right ) \, .
\end{array}
\right . 
\end{equation}
Notice that, differently from the case of $J_{n2}$, the contribution of $J_{02} (\alpha)$ starts at $ {\cal O} (1/\alpha)$. Eventually, 
\begin{equation}
I_0^+ (\alpha) = \frac{1}{2} \ln \frac{\pi}{2 \alpha} - \frac{1}{\alpha} +  \frac{3}{4 \alpha^2} + {\cal O} \left (\frac{1}{\alpha^3} \right ) \, .
\end{equation}
The positive would-be zero-mode energy has, therefore, the following form in the strong coupling limit: 
\begin{equation}
\label{eq:k0strongcouplingp}
k_0^+(\alpha) \simeq \frac{1}{2R} \exp \left [-\frac{1}{\alpha} +  \frac{3}{4 \alpha^2} + {\cal O} \left (\frac{1}{\alpha^3} \right ) \right ] \, .
\end{equation}
Notice that the strong coupling limit of $k_{0^+}$ and $k_n$ (with $n \geq 1$) is identical at first order in $1/\alpha$. The difference in the ${\cal O}(1/\alpha^2)$
term between the $n=0$ and the $n \geq 1$ eigenvalues is relevant to compute the contribution of $\psi_{0^+}^e$ to the Casimir energy, as I will show in a
forthcoming publication.

For negative $\alpha$, the strong coupling expansion of the would-be zero-mode energy can be carried out in a similar way: 
\begin{equation}
k^-_0 = \frac{i}{R}\sqrt{\frac{|\alpha|}{ 2 \pi}} \, \exp \left [ I^-_0 (-|\alpha|) \right ] = \frac{1}{2 R} \exp \left (\frac{1}{2} \ln \frac{2 \alpha}{\pi} + I^-_0 (-|\alpha|) \right ) \, ,
\end{equation}
where
\begin{equation}
I^-_0(-|\alpha|) = -\frac{1}{\pi} \int_0^1 dt \frac{1}{t} \left \{ \arctan \left ( \frac{\pi t}{2 |\alpha| \lambda} \right ) - \frac{\pi}{2}  \right \} \, .
\end{equation}

Changing variables, $t \to (2 |\alpha|/\pi) \, x$, we get:
\begin{equation}
I^-_0(-|\alpha|) = - \frac{1}{\pi} \int_0^{2|\alpha|/\pi} dx \frac{1}{x} \, g^-(x) = 
- \left . \frac{1}{\pi}   \ln x g^-(x)  \right |_{0}^{2|\alpha|/\pi} + \frac{1}{\pi} \int_0^{2 |\alpha|/\pi} dx \ln x \frac{d g^-(x)}{dx} \, ,
\end{equation}
where
\begin{equation}
g^-(x) = - \frac{\pi}{2} + \left . \arctan \left ( \frac{\pi t}{2 |\alpha| \lambda} \right ) \right |_{t = (\pi/ 2|\alpha|) x} \, .
\end{equation}
Expanding in powers of $1/|\alpha|$:
\begin{equation}
I_0^-(-|\alpha|) = A^-_0 (|\alpha|) + J^-_{01}(|\alpha|) +  J^-_{02}(|\alpha|)  +  J^-_{03}(|\alpha|) + \dots  \, ,
\end{equation}
where (up to second order in $1/|\alpha|$):
\begin{equation}
\left \{
\begin{array}{lll}
A^-_0 (|\alpha|) &=& -\left . \frac{1}{\pi}  \ln x g^-(x) \right |_{0}^{2|\alpha|/\pi} = -\frac{1}{2} \ln \frac{\pi}{2|\alpha|}  + \frac{1}{2|\alpha|}  \ln \frac{\pi}{2|\alpha|} 
+\frac{1}{2 |\alpha|^2} \ln \frac{\pi}{2 |\alpha|}
+ {\cal O} \left (\frac{1}{|\alpha|^3} \right ) \, , \\
\\
J^-_{01} (|\alpha|) &=& - J^+_{01} (|\alpha|) \, , \\
\\
J^-_{02} (|\alpha|) &=&  J^+_{02} (|\alpha|)  \, , \\
\\
J^-_{03} (|\alpha|) &=& - J^+_{03} (|\alpha|)  \, .
\end{array}
\right . 
\end{equation}
Eventually, 
\begin{equation}
I_0^- (-|\alpha|) = -\frac{1}{2} \ln \frac{\pi}{2 |\alpha|} +  \frac{1}{4 |\alpha|^2} + {\cal O} \left (\frac{1}{|\alpha|^3} \right ) \, .
\end{equation}
The negative would-be zero-mode energy has, therefore, the following form: 
\begin{equation}
\label{eq:k0strongcouplingm}
k_0^-(-|\alpha|) \simeq \frac{i |\alpha|}{ \pi R} \exp \left [\frac{1}{4 |\alpha|^2} + {\cal O} \left (\frac{1}{|\alpha|^3} \right ) \right ] \, ,
\end{equation}
{\em i.e.} it grows linearly with $|\alpha|$.

For $|\alpha| \to \infty$ and $n>0$, the eigenmodes become:
\begin{equation}
\lim_{q R \to \infty} \psi_n^e (y) = \frac{1 }{\sqrt{\pi R}} \cos  \left [ \left ( n + \frac{1}{2} \right ) \frac{(|y|- \pi R)}{R} \right ] 
\propto \frac{1}{\sqrt{\pi R}} \sin  \left [ \left ( n + \frac{1}{2} \right ) \frac{|y|}{R} \right ]
\end{equation}
after rotating away an unphysical phase factor $(-1)^n$.

\section{Adding relevant operators to the action}
\label{sec:interactions}

I will now add two relevant operators (according to the dimensional classification given in Tab.~\ref{tab:operators}) to eq.~(\ref{eq:scalarplusbraneaction}):
\begin{equation}
\label{eq:scalarplusbraneaction2}
S = \int d^4x \int_{-\pi R}^{\pi R} dy \left \{ \frac{1}{2} \partial^M \Phi(x_\mu,y) \partial_M \Phi(x_\mu,y)  - \frac{q}{2} \, \delta (y) \, \Phi^2(x_\mu,y)
- \frac{m^2}{2}  \, \Phi^2(x_\mu,y) - \frac{\lambda}{3!}  \Phi^3(x_\mu,y) \right \}  \, .
\end{equation}
The first term added to the action studied in the previous sections is a bulk mass term for the scalar field $\Phi (x_\mu, y)$. This term does not modify 
the previous results: after decomposing the 5-dimensional massive equation of motion
\begin{equation}
\label{eq:branescalarmass}
\Box_5 \Phi(x_\mu,y) + m^2 \,  \Phi (x_\mu, y) + q \, \delta(y) \Phi(x_\mu,y) =  0 \, .
\end{equation}
into two separate equations:
\begin{equation}
\label{eq:branescalarmass2}
\left \{
\begin{array}{l}
\Box_4 \phi(x_\mu) + m^2 \phi (x_\mu) = -\lambda^2 \phi(x_\mu) \, , \\
\\
\frac{d^2}{dy^2} \psi(y) - q \, \delta(y) \psi(y) = -\lambda^2 \psi(y) \, ,
\end{array}
\right .
\end{equation}
we get the same results as before with the only difference that the 4-momentum $p_\mu$ satisfies the relation $p_\mu p^\mu = \lambda^2 - m^2$
and that, therefore, the resulting dispersion relation is: 
\begin{equation}
\label{eq:massivedispersionrelation}
\omega_n = \sqrt{m^2 +|\vec{p}|^2 + k_n^2} \, ,
\end{equation}
where $k_n$ are the eigenvalues given in Sects.~\ref{sec:oddmodes} and \ref{sec:evenmodes}.

The cubic operator $\Phi^3 (x_\mu, y)$, on the other hand, is the only relevant interaction that can be added to the lagrangian without spoiling its superficial renormalizability. 
The coupling $\lambda$, in 5 dimensions, has dimension $1/2$. In principle, it is possible to parametrize it as a function of the 
brane tension $f$, $\lambda = c^\prime \, f^{1/2}$ (as for the case of the brane-to-bulk coupling). 
However, as we have already introduced a bulk mass term $m$ for the scalar field $\Phi$, it is no longer true that $f$ is the only
physical scale present in the theory\footnote{It is reasonable, however, to think that $f \gg m$, as we need a stiff brane to avoid that brane-localized modes do not fill the extra-dimension moving across the bulk with the brane fluctuations. From this point of view, therefore, $m$ would be a scale that affects the infra-red properties of the theory, whereas its
ultra-violet properties are governed by the scale $f$ \cite{delAguila:2003bh,delAguila:2006atw}.}. 
The cubic interaction of the bulk field $\Phi$ can be decomposed into interactions of the even and odd 
4-dimensional Kaluza-Klein modes. There are four possible couplings between KK-modes: two trivially vanishing ones: $g^{eeo}_{lmn} = g^{ooo}_{lmn} = 0$;
and two non-trivial ones, 
\begin{equation}
\left \{
\begin{array}{lll}
g^{eoo}_{lmn} & = & \lambda \int_{- \pi R}^{\pi R} dy \,  \psi^e_l (y) \, \psi^o_m (y) \, \psi^o_n (y) \qquad 
\left \{ 
\begin{array}{lll}
l &=& 0, 1, 2, \dots \\
 m,n &=& 1,2, ,\dots
 \end{array}
\right .
\\
\\
g^{eee}_{lmn} & = & \lambda \int_{- \pi R}^{\pi R} dy \,  \psi^e_l (y) \, \psi^e_m (y) \, \psi^e_n (y) \qquad l, m, n = 0, 1, 2, \dots 
\end{array}
\right .
\end{equation}
Using eqs.~(\ref{eq:oddmodes}) and (\ref{eq:evenmodes}) I get: 
\begin{equation}
\left \{
\begin{array}{lll}
g^{eoo}_{lmn}  (\alpha) &=  \frac{\lambda}{\sqrt{\pi R}}  \, c_{lmn} (\alpha) &= \frac{\lambda}{\sqrt{\pi R}} \times h_l (\alpha) \,  \int_{-1}^1 da \, \cos \left [ \xi_l (\alpha) (|a|-1)\right ] \, \sin \pi m a \, \sin \pi n a \, , \\
\\
g^{eee}_{lmn}(\alpha) &=  \frac{\lambda}{\sqrt{\pi R}}  \, d_{lmn} (\alpha ) &= \frac{\lambda}{\sqrt{\pi R}} \times  h_l (\alpha) \,  h_m (\alpha)  \,  h_n (\alpha) \,  \\
&& \times  \int_{-1}^1 da \, \cos \left [ \xi_l (\alpha) (|a|-1)\right ]   \, \cos \left [ \xi_m (\alpha) (|a|-1)\right ] \, \cos  \left [ \xi_n (\alpha) (|a|-1)\right ] \, , \\
\end{array}
\right .
\end{equation}
where $c_{lmn}, d_{lmn}$ are dimensionless $\alpha$-dependent coefficients and $\lambda/\sqrt{\pi R}$ represents the (dimension 1) physical 4-dimensional 
parameter that defines the strength of the interaction between three 4-dimensional scalar fields.
Notice that these formul\ae \, apply to $l,m,n \geq 1$ for any value of $\alpha$ and to $l,m$ or $n=0$ for $\alpha$ positive. 
In the case of $l=0$, $\alpha$ negative, we have instead: 
\begin{equation}
\left \{
\begin{array}{lll}
g^{eoo}_{0^-mn}  (\alpha ) &=  \frac{\lambda}{\sqrt{\pi R}}  \, c_{0^-mn} (\alpha) &= \frac{\lambda}{\sqrt{\pi R}} \times h_{0^-} (\alpha) \,  \int_{-1}^1 da \, \cosh \left [ |\xi_0^- (\alpha) |(|a|-1)\right ] \, \sin \pi m a \, \sin \pi n a \, , \\
\\
g^{eee}_{0^-mn}(\alpha) &=  \frac{\lambda}{\sqrt{\pi R}}  \, d_{0^-mn} (\alpha ) &= \frac{\lambda}{\sqrt{\pi R}} \times  h_{0^-} (\alpha) \,  h_m (\alpha)  \,  h_n (\alpha) \,  \\
&& \times \int_{-1}^1 da \, \cosh \left [ |\xi_0^-(\alpha)| (|a|-1)\right ]   \, \cos \left [ \xi_m (\alpha) (|a|-1)\right ] \, \cos  \left [ \xi_n (\alpha) (|a|-1)\right ]\, ,  \\
\\
g^{eee}_{0^-0^-n}(\alpha) &=  \frac{\lambda}{\sqrt{\pi R}}  \, d_{0^-0^-n} (\alpha ) &= \frac{\lambda}{\sqrt{\pi R}} \times  h^2_{0^-} (\alpha) \,   h_n (\alpha) \,  \\
&& \times \int_{-1}^1 da \, \cosh \left [ |\xi_0^-(\alpha)| (|a|-1)\right ]   \, \cosh \left [ |\xi_0^- (\alpha)| (|a|-1)\right ] \, \cos  \left [ \xi_n (\alpha) (|a|-1)\right ] \, , \\
\\
g^{eee}_{0^-0^-0^-}(\alpha) &=  \frac{\lambda}{\sqrt{\pi R}}  \, d_{0^-0^-0^-} (\alpha ) &= \frac{\lambda}{\sqrt{\pi R}} \times  h^3_{0^-} (\alpha) \\
&& \times \int_{-1}^1 da \, \cosh \left [ |\xi_0^-(\alpha)| (|a|-1)\right ]   \, \cosh \left [ |\xi_0^- (\alpha)| (|a|-1)\right ] \, \cosh  \left [ |\xi_0^- (\alpha)| (|a|-1)\right ] \, , \\
\end{array}
\right .
\end{equation}
where $m,n \geq 1$.

These couplings can be computed numerically using the expressions for the energy of the Kaluza-Klein modes, $\xi^\pm_0 (\alpha)$ and $\xi_n (\alpha)$,
given in previous sections.  Of particular interest are the couplings of the would-be zero-modes  (either $\psi_{0^+}^e$ or $\psi_{0^-}^e$) 
with the Kaluza-Klein modes of higher KK-number. These will be given below separately for the case of $\alpha > 0$ and $\alpha < 0$. 
 
 \subsection{Couplings of the positive would-be zero-mode $\psi_{0^+}^e$ with higher KK-modes}
 
The analytic expression for the $eoo$-coupling is rather simple:
\begin{equation}
c_{0^+mn} = -  \frac{4 \pi^2 m n \, \xi_0^+(\alpha) \, \sin \xi_0^+ (\alpha)}{\left \{ 
\left [ \xi_0^+ (\alpha) \right ]^4  - 2 \pi^2 (m^2 + n^2) \, \left [ \xi_0^+ (\alpha) \right ]^2 + \pi^4 (m^2 - n^2)^2 \right \} 
} \;  h_{0^+} (\alpha) \qquad \alpha > 0 \, ,
\end{equation}
for $m,n \geq 1$. 

On the other hand, the expression for the $eee$-coupling is not so simple: 
\begin{eqnarray}
d_{0^+mn} & = &-  2 \frac{h_{0^+} h_m h_n}{\left [\xi_0^{+2}-(\xi_m - \xi_n)^2 \right  ] \left [ \xi_0^{+2}-(\xi_m + \xi_n)^2 \right ]} \nonumber \\
& \times & \left \{ \cos \xi_m \left [  \xi_0^+ \left ( \xi_m^2 + \xi_n^2 -\xi_0^{+2}  \right ) \cos \xi_n \sin \xi_0^+ +
                                                        \xi_n \left (  \xi_m^2 - \xi_n^2 + \xi_0^{+2}\right ) \sin \xi_n \cos \xi_0^+ \right ]  \right . \nonumber \\
&+& \left . \xi_m \sin \xi_m \left [  \xi_0^+ \left (- \xi_m^2 + \xi_n^2  + \xi_0^{+2} \right ) \cos \xi_n \cos \xi_0^+ 
+ 2 \xi_0^+ \xi_n \sin \xi_n \sin \xi_0^+ \right ]
\right \}
\end{eqnarray}

\begin{figure}[htb!]
\begin{flushleft}
	\includegraphics[width=16cm]{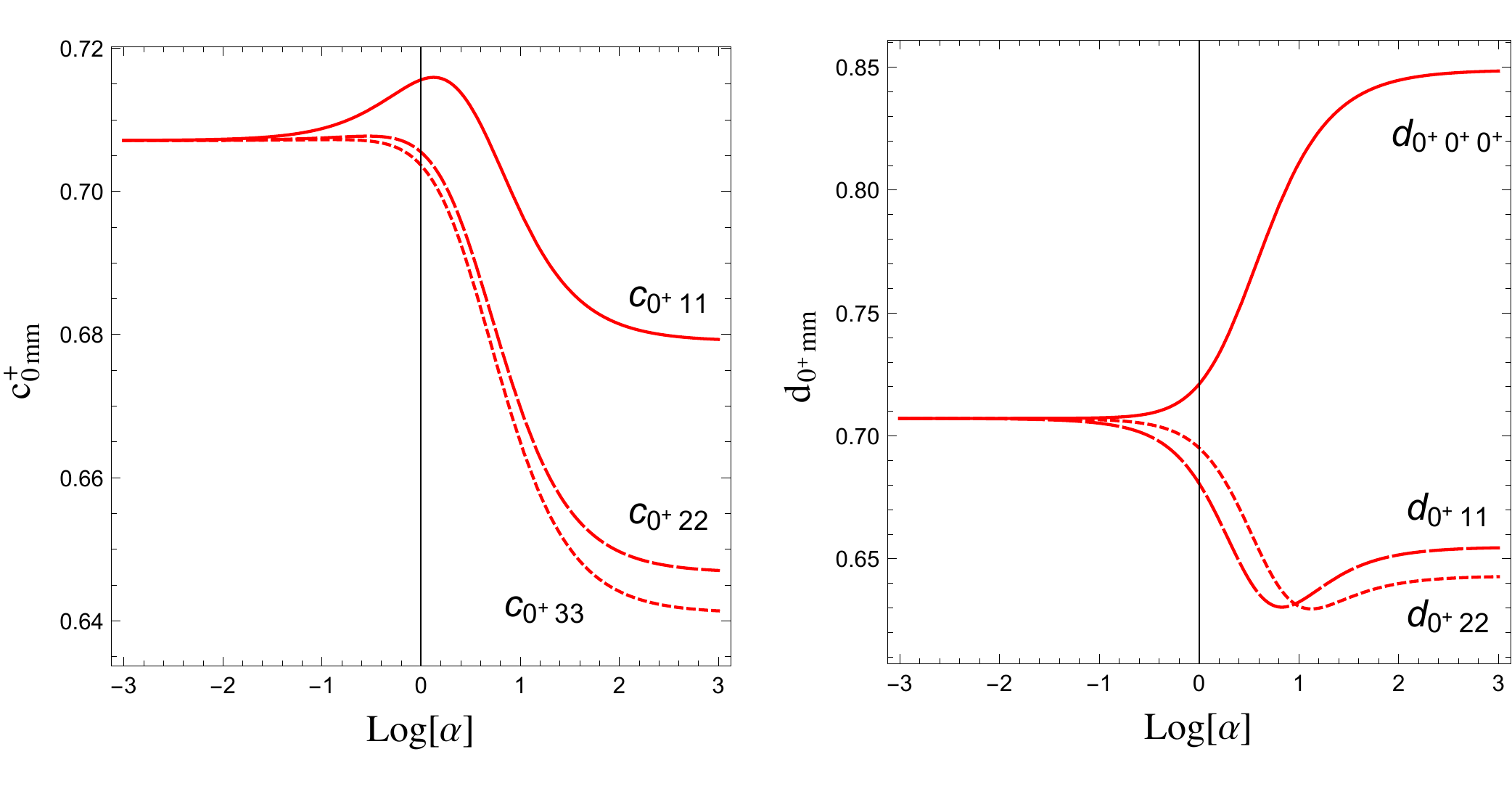}
\caption{\it The $\alpha$-dependence of diagonal cubic couplings of the would-be zero-mode $\psi_{0^+}^e (\alpha)$ for positive $\alpha$.
Left panel: the coupling with two odd modes, $c_{0^+mm}(\alpha)$, for $m = 1,2,3$; right panel: the coupling with two even modes, $d_{0^+mm} (\alpha)$, for $m = 0,1,2$. }
\label{fig:diagonalgpos}
\end{flushleft}
\end{figure}

The behaviour of the diagonal $eoo$ and $eee$ couplings ($c_{0^+mm}$ and $d_{0^+mm}$) as a function of $\alpha$, is depticted in Fig.~\ref{fig:diagonalgpos}.
The left panel shows the $g^{eoo}$-coupling of $\psi_{0^+}^e$ with the lightest odd KK-modes, $m = 1,2,3$ (solid, dashed and dotted red lines, respectively). The right panel shows the $g^{eee}$-coupling with the lightest even KK-modes for $m = 1,2$ (solid, dashed and dotted red lines, respectively). 
The cubic self-interaction of $\psi_{0^+}^e$, $d_{0^+0^+0^+}$, is also shown. In both panels we can see that, for small $\alpha$,  the universal coupling 
$c_{0^+mm} = d_{0^+mm} = 1/\sqrt{2}$ is reached. Remember that in the limit of vanishing $\alpha$ the brane is absent and we should recover the 
standard KK-decomposition of a cubic self-interaction of the 5-dimensional field $\Phi$. In particular, in the limit $\alpha \to 0$ only diagonal couplings of $\psi_{0^+}^e$ 
with the massive KK-modes are allowed, as momentum in the 5th-dimension is conserved. 
For large $\alpha$, on the other hand, universality is broken and the interaction between $\psi_{0^+}^e$ and massive KK-modes
approaches exponentially fast $n$-dependent asymptotical values. Analytic expressions for the large $\alpha$ limit, obtained using eqs.~(\ref{eq:knstrongcoupling})
and (\ref{eq:k0strongcouplingp}), valid for KK-modes such that $m/ \alpha \gg 1$, are given below:
\begin{eqnarray}
c_{0^+mm} &=& \frac{32}{\pi} \frac{m^2}{16 m^2-1} \left \{ 1- \frac{1}{\alpha} \left [ \frac{3}{2} - 4 \frac{4 m^2 -1}{16 m^2 -1} \right ] 
+ {\cal O} \left ( \frac{1}{\alpha^2} \right )  \right \}  \, ,  \\
\nonumber \\
d_{0^+mm} &=& \frac{8}{\pi} \frac{(2 m +1)^2}{(4 m +1) (4 m +3)} \left [ 1 - \frac{1}{2 \alpha} 
+ {\cal O} \left ( \frac{1}{\alpha^2} \right )     \right ] \, . 
\end{eqnarray}
The expression for the strong coupling limit of $c_{0^+mm}$ is valid for $m \geq 1$ whereas, at the leading order in $1/\alpha$ the expression for $d_{0^+mm}$ is valid for 
$m \geq 0$ ($m$-dependent corrections arise at order $1/\alpha^2$). For $m \to \infty$, we get $c_{0^+mm} = d_{0^+mm} = 2/ \pi \left [1 - 1/(2 \alpha) \right ] + {\cal O}(1/m^2)$.

\begin{figure}[htb!]
\begin{flushleft}
	\includegraphics[width=16cm]{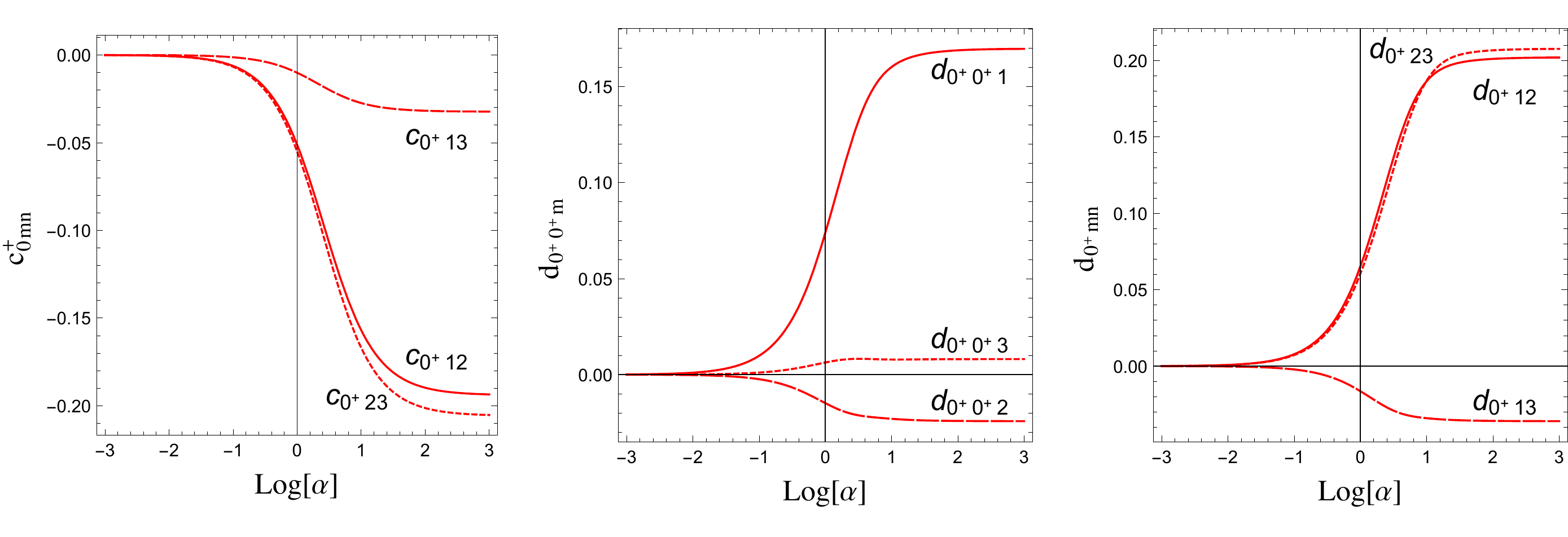}
\caption{\it The $\alpha$-dependence of non-diagonal cubic couplings of the would-be zero-mode $\psi_{0^+}^e (\alpha)$ for positive $\alpha$.
Left panel: the coupling with two odd modes, $c_{0^+mn}(\alpha)$, for $(m,n) = (1,2),(1,3),(2,3)$; 
middle panel: the coupling of two zero-modes with one even mode, $d_{0^+0^+m} (\alpha)$, for $m=1,2,3$;
right panel: the coupling with two even modes, $d_{0^+mn} (\alpha)$, for $(m,n) = (1,2),(1,3),(2,3)$.
}
\label{fig:nondiagonalgpos}
\end{flushleft}
\end{figure}

An important consequence of the coupling between the bulk field $\Phi$ and the brane is that non-diagonal interactions are no longer vanishing, as momentum
conservation in the 5th-dimension is explicitly broken by the presence of the brane. The behaviour of the non-diagonal $eoo$ and $eee$ couplings for positive $\alpha$, 
as a function of $\alpha$, is depticted in Fig.~\ref{fig:nondiagonalgpos}.
The left panel shows the $g^{eoo}$-coupling of the would-be zero-mode $\psi_{0^+}^e$ with the lightest odd KK-modes, $(m,n) = (1,2),(1,3),(2,3)$ 
(solid, dashed and dotted red lines, respectively). The middle panel shows the non-diagonal $g^{eee}$ coupling between two would-be zero-modes
and a massive even KK-mode, $m = 1,2,3$ (solid, dashed and dotted red lines, respectively). Eventually, the right panel shows the non-diagonal
$g^{eee}$ coupling between one would-be zero-mode and two massive even KK-modes, $(m,n) = (1,2),(1,3),(2,3)$ (solid, dashed and dotted red lines, respectively).
In this case, we can clearly see that all non-diagonal couplings vanish in the limit $\alpha \to 0$, pointing to restoration of momentum conservation in the 5th-dimension.
Non-diagonal, non-vanishing, couplings are instead allowed in the limit $\alpha \to \infty$.

The large $\alpha$ expansion for non-diagonal couplings (valid for $m/\alpha, n/\alpha \gg 1$) is: 
\begin{eqnarray}
c_{0^+mn} & = & - \frac{32}{\pi} \frac{m n}{16 (m^2-n^2)^2 - 8 (m^2+n^2)+1} \left \{ 
1 - \frac{1}{\alpha} \left [
\frac{3}{2} + 4 \frac{4 (m^2 + n^2)-1}{16 (m^2-n^2)^2 - 8 (m^2+n^2)+1} \right ] 
+ {\cal O}\left ( \frac{1}{\alpha^2} \right )  \right \} \, , \nonumber \\
\\
d_{0^+mn} & = & (-1)^{m+n+1} \frac{8}{\pi} \frac{(2 m +1) (2 n +1)}{(2 m + 2 n +1) (2 m - 2 n +1) (2 m - 2 n -1) (2 m + 2 n + 3)} \left \{ 1 - \frac{1}{2 \alpha}  
+ {\cal O} \left ( \frac{1}{\alpha^2}\right ) \right \} \, .  \nonumber \\
\end{eqnarray}

From these equations the diagonal limit is easily recovered putting $m = n$. On the other hand, it is interesting to see that the non-diagonal couplings vanish
in the limit of fixed $m$ and large $n$. For example, $d_{0^+0^+n} = {\cal O} (1/n^3)$ for $n \to \infty$. This behaviour is compatible with what found in the
case of BLKT in Refs.~\cite{Dvali:2001gm,Dvali:2001gx,Carena:2002me}.

\subsection{Couplings of the negative localized would-be zero-mode $\psi_{0^-}^e$ with higher KK-modes}

Far more interesting phenomenologically is the case of the localized would-be zero-mode, $\psi_{0^-}^e$. 
In this case I get for the $g^{eoo}$ coupling:
\begin{equation}
c_{0^-mn} =   \frac{4 \pi^2 m n \, \xi_0^-(\alpha) \, \sinh \xi_0^- (\alpha)}{\left \{ 
\left [ \xi_0^- (\alpha) \right ]^4  + 2 \pi^2 (m^2 + n^2) \, \left [ \xi_0^- (\alpha) \right ]^2 + \pi^4 (m^2 - n^2)^2 \right \} 
} \;  h_{0^-} (\alpha) \qquad \alpha < 0 \, ,
\end{equation}
valid for $m, n \geq 1$. For the $g^{eee}$ coupling  the number of zero-modes involved must be specified. I get:
\begin{equation}
\left \{
\begin{array}{lll}
d_{0^-mn} & = &-  2 \frac{h_{0^-} h_m h_n}{\left [\xi_0^{-2} + (\xi_m - \xi_n)^2 \right  ] \left [ \xi_0^{-2} + (\xi_m + \xi_n)^2 \right ]} \\
& \times & \left \{ \cosh \xi_0^- \left [  \xi_m \left ( \xi_m^2 - \xi_n^2  + \xi_0^{-2}  \right ) \cos \xi_n \sin \xi_m -
                                                        \xi_n \left (  \xi_m^2 - \xi_n^2 - \xi_0^{-2}\right ) \sin \xi_n \cos \xi_m \right ]  \right .  \\
&+& \left . \xi_0^- \sinh \xi_0^- \left [  \left ( \xi_m^2 + \xi_n^2  + \xi_0^{-2} \right ) \cos \xi_m \cos \xi_n 
+ 2 \xi_m \xi_n \sin \xi_m \sin \xi_n \right ]
\right \} \qquad \qquad (m, n \geq 1) \\
\\
d_{0^-0^-n} & = & \frac{h_{0^-}^2 h_n}{\xi_n \left ( 4 \xi_0^{-2} + \xi_n^2 \right ) } \times \left \{
\left ( 4 \xi_0^{-2} + \xi_n^2 \right ) \sin \xi_n + \xi_n^2 \cosh 2 \xi_0^- \sin \xi_n + 2 \xi_0^- \xi_n \sinh 2 \xi_0^- \cos \xi_n  \right \} \;  (n \geq 1) \\
\\
d_{0^-0^-0^-} &=& 6 \frac{h_{0^-}^3}{\xi_0^-} \times \left \{ 9 \sinh \xi_0^- + \sinh 3 \xi_0^- \right \}
\end{array}
\right .
\end{equation}

\begin{figure}[htb!]
\begin{flushleft}
	\includegraphics[width=16cm]{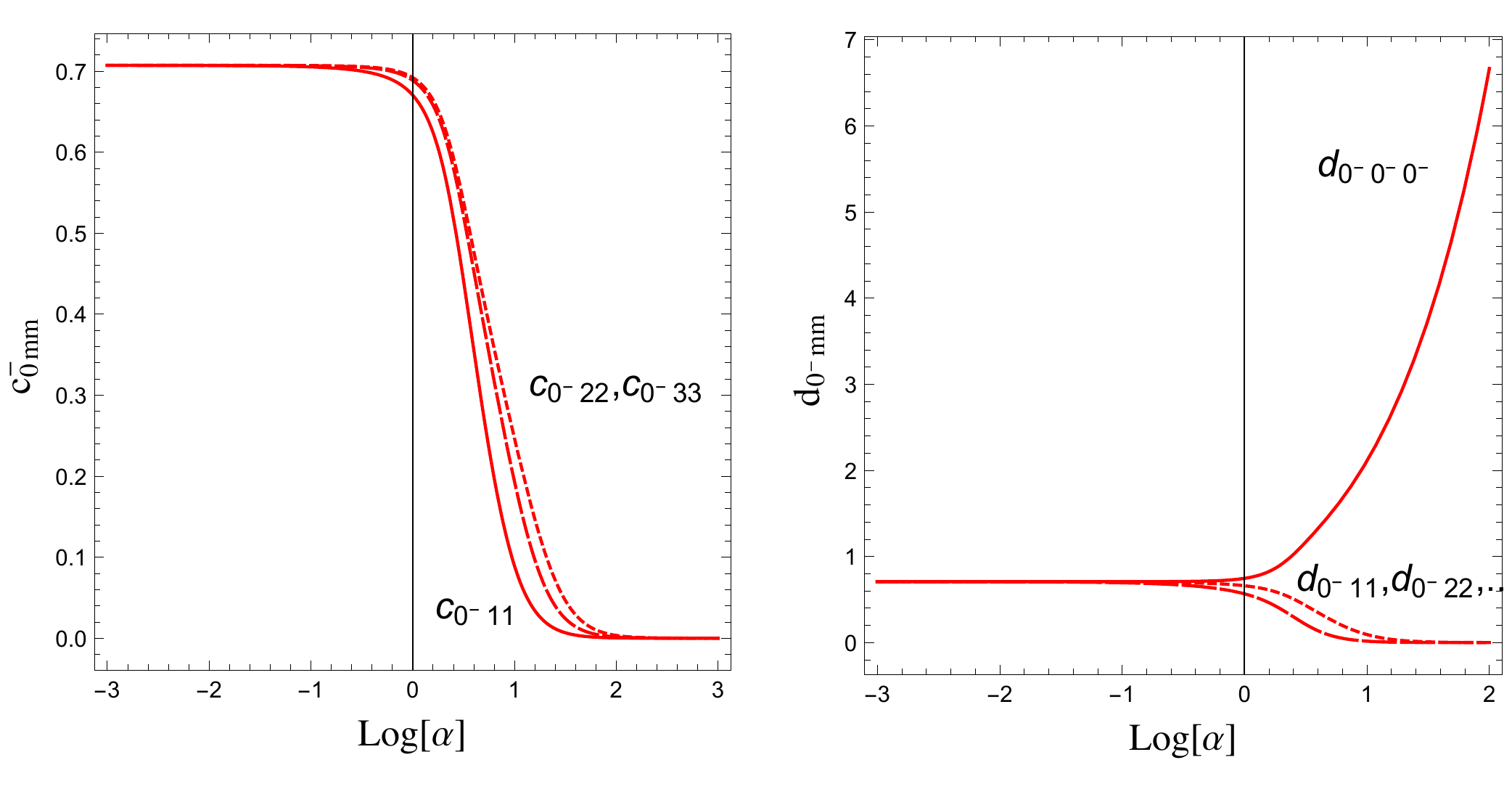}
\caption{\it The $\alpha$-dependence of diagonal cubic couplings of the localized would-be zero-mode $\psi_{0^-}^e (\alpha)$ for negative $\alpha$.
Left panel: the coupling with two odd modes, $c_{0^-mm}(\alpha)$, for $m = 1,2,3$; right panel: the coupling with two even modes, $d_{0^-mm} (\alpha)$, for $m = 0,1,2$. }
\label{fig:diagonalgneg}
\end{flushleft}
\end{figure}

The behaviour of the diagonal $eoo$ and $eee$ $\psi_{0^-}^e$ couplings $c_{0^-mm}$ and $d_{0^-mm}$, as a function of $\alpha$, is depticted in Fig.~\ref{fig:diagonalgneg}.
The left panel shows the $g^{eoo}$-coupling of the localized would-be zero-mode $\psi_{0^-}^e$ with the lightest odd KK-modes, $m = 1,2,3$ (solid, dashed and dotted red lines, respectively). The right panel shows its $g^{eee}$-coupling with the lightest even KK-modes for $m = 1,2$ (solid, dashed and dotted red lines, respectively). 
The cubic self-interaction of $\psi_{0^-}^e$, $d_{0^-0^-0^-}$ is also shown. As it was the case for $\alpha > 0$, for small $\alpha$ universality is recovered and
$c_{0^-mm} = d_{0^-mm} = 1/\sqrt{2}$ is reached in the limit $\alpha \to 0$. On the other hand, for large $\alpha$ all diagonal couplings 
with $m \geq 1$ vanish exponentially fast. Their expression in the strong coupling limit (for $m/\alpha \gg 1$) are:
\begin{eqnarray}
c_{0^-mm} &=& 4 \pi^2 \frac{m^2}{\alpha^{5/2}} \, \left [ 1 + {\cal O} \left ( \frac{1}{\alpha^2} \right )  \right ]  \, ,  \\
\nonumber \\
d_{0^-mm} &=& \frac{\pi^2}{2} \frac{(2 m -1)^2}{\alpha^{5/2}} \, \left [ 1 + \frac{3}{4 \alpha}  + {\cal O} \left ( \frac{1}{\alpha^2} \right )     \right ] \, .
\end{eqnarray}
The cubic self-coupling of the localized would-be zero-mode, $d_{0^-0^-0^-}$, has a different behaviour:
\begin{equation}
d_{0^-0^-0^-} =  \frac{2}{3} \alpha^{1/2} \left [1 +  {\cal O} \left ( \frac{1}{\alpha^2} \right )  \right ] \, ,
\end{equation}
as this is the only coupling that survives at tree-level in the limit of infinite $\alpha$.  

\begin{figure}[htb!]
\begin{flushleft}
	\includegraphics[width=16cm]{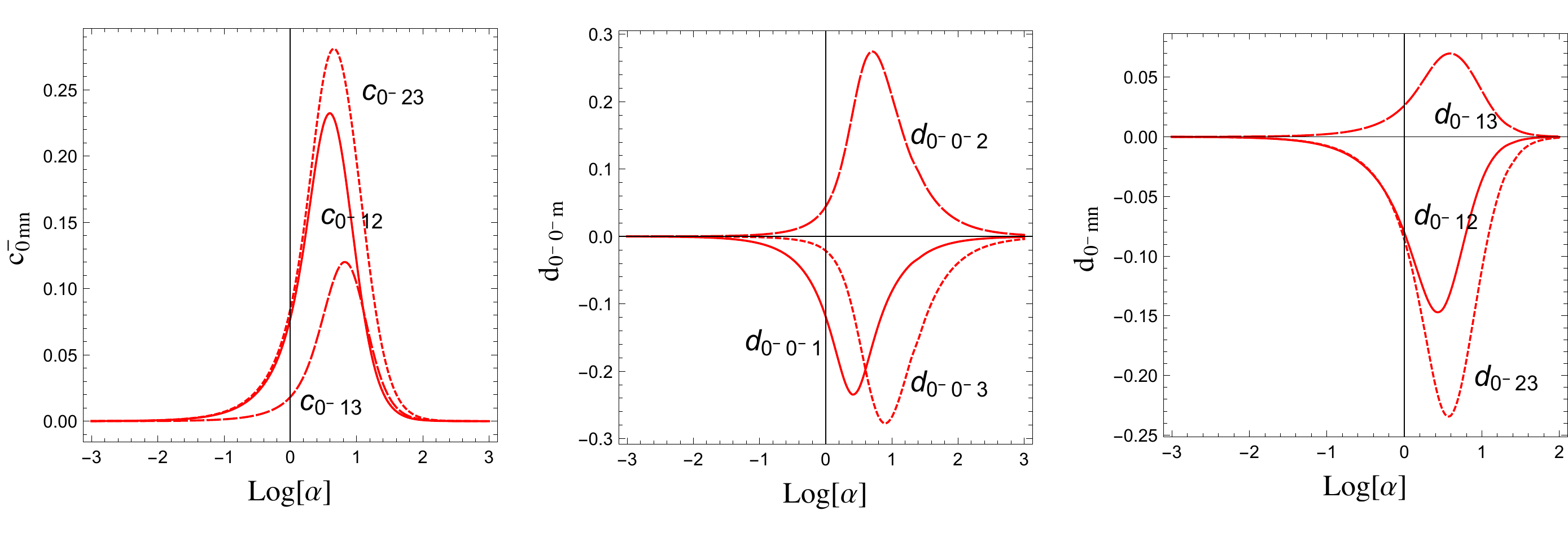}
\caption{\it The $\alpha$-dependence of non-diagonal cubic couplings of the localized would-be zero-mode $\psi_{0^-}^e (\alpha)$ for negative $\alpha$.
Left panel: the coupling with two odd modes, $c_{0^-mn}(\alpha)$, for $(m,n) = (1,2),(1,3),(2,3)$; 
middle panel: the coupling of two zero-modes with one even mode, $d_{0^-0^-m} (\alpha)$, for $m=1,2,3$;
right panel: the coupling with two even modes, $d_{0^-mn} (\alpha)$, for $(m,n) = (1,2),(1,3),(2,3)$.
}
\label{fig:nondiagonalgneg}
\end{flushleft}
\end{figure}

The non-diagonal couplings $c_{0^-mn}$ and $d_{0^-mn}$ are shown in Fig.~\ref{fig:nondiagonalgneg}. The left panel shows the non-diagonal $c_{0^-mn}$ couplings
for $(m,n) = (1,2),(1,3),(2,3)$ (solid, dashed and dotted red lines, respectively). The middle panel shows the $d_{0^-0^-m}$ coupling with $m =1,2,3$ (solid, dashed and dotted 
red lines, respectively). Eventually, the right panel shows the $d_{0^-mn}$ couplings for $(m,n) = (1,2),(1,3),(2,3)$ (solid, dashed and dotted red lines, respectively).
In all panels, it can be seen that all non-diagonal couplings vanish for $\alpha \to 0$, as it is expected due to momentum conservation in the 5th-dimension. 
For intermediate $\alpha$, $\alpha \in [10^{-1},10^1]$, the non-diagonal couplings can become rather large (values of the couplings as large as $0.3$ can be found
for all the non-diagonal couplings). Analytical expressions are not much inspiring in this regime. However, we can see that all couplings also vanish in the strong coupling limit. 
Their strong coupling expressions (valid for $m/\alpha, n/\alpha \gg1$) are given below: 
\begin{eqnarray}
c_{0^-mn} & = & 4 \pi^2 \frac{m n}{\alpha^{5/2}} \, \left [ 1 + {\cal O} \left ( \frac{1}{\alpha^2} \right )  \right ]  \, ,  \\
\nonumber \\
d_{0^-0^-m} &=& (-1)^n (2 m -1) \frac{\pi}{4 \alpha} \left [ 1 + \frac{3}{2 \alpha} + {\cal O} \left ( \frac{1}{\alpha^2} \right ) \right ] \, , \\
\nonumber \\
d_{0^-mn} & = & (-1)^{m+n} \frac{\pi^2}{2 \alpha^{5/2}} (2 m - 1) (2 n -1) \left \{ 1 + \frac{3}{4 \alpha}  + {\cal O} \left ( \frac{1}{\alpha^2}\right ) \right \} \, . 
\end{eqnarray}
From these expressions we see that all non-diagonal couplings vanish as ${\cal O }(1/\alpha^{5/2})$, with the exception of $d_{0^-0^-m}$ whose approach to zero is 
only ${\cal O}(1/\alpha)$.
A straightforward consequence is that, in the strong coupling limit, the localized would-be zero-mode $\psi_{0^-}^e$ becomes effectively decoupled from all the KK-modes. 
For $\alpha \to \infty$ only the cubic self-coupling $d_{0^-0^-0^-}$ survives and the classical action of the 4-dimensional $\psi_{0^-}^e$ field lose any knowledge
about the details of the fundamental 5-dimensional theory. On the other hand, for large but non-infinite $\alpha$, the localized would-be zero-mode becomes sensitive
to quantum corrections involving all the KK-modes (both even and odd) present at the fundamental level. 

\section{Conclusions}
\label{sec:concl}

The existence of spatial dimensions that exceed the three observed at ordinary human-size scales in all physics phenomena up to date was proposed
long ago by T. Kaluza and O. Klein.  The idea was so appealing that A. Einstein tried (unsuccessfully) for long to use it to build a non-quantum unified theory of general relativity and electromagnetism. As recalled in the Introduction, albeit abandoned to this purpose, the idea was retaken at the end of the '90s
to explain the large hierarchy existing between the Planck scale ($M_P \sim 10^{19}$ GeV) and the electro-weak symmetry breaking scale 
($\Lambda_{\rm EW} \sim 250$ GeV).
Several options involving extra spatial dimensions have been advanced since then. The one in which I am interested in this paper is the so-called 
 {\em Large Extra-Dimensions} model (LED). In this model, in order to escape observation in present experiments, extra-dimensions are compactified 
 in a volume $V_n$ with sub-mm size. To solve the hierarchy problem, however, they must be "large" enough that $M_P^2 \sim V_n \times M_D^{n+2}$ 
 can give a small hierarchy between the fundamental, $(4+n)$-dimensional, scale of gravity  $M_D$ and $\Lambda_{\rm EW}$ (thus motivating the name of the model).
  
A necessary ingredient in a model that goes Beyond the Standard Model is to include the Standard Model itself as a low-energy effective theory 
recovered when the energy scale testable at experiments is (much) lower than the typical scales of the model. The way to introduce Standard Model fields and particles 
into the LED scenario is through the concept, borrowed from string theory, of $D$-branes, {\em i.e.} the locus in space-time where the ending points (representing gauge fields) 
of an open string are bounded by Dirichlet boundary conditions. Maybe overusing this idea, not only SM gauge fields are supposed to be confined to the brane in the LED model, but all SM fields charged under the SM gauge group $SU_c(3) \times SU_L(2)\times U_Y(1)$, with the possible exception of right-handed neutrinos that are neutral under it.
After compactification of the extra-dimensions no Kaluza-Klein towers arise in correspondence of brane-localized fields 
(thus explaining the non-observation of light, but massive, photons). 
In the standard approach, the brane is introduced by hand adding to the gravitational action
$S_{\rm EH}$ a term $S_{\rm brane} = \int d^4 x \int_{V_n} d^n y \, \delta^{(n)} (y) \, {\cal L}_{\rm SM} (x,y)$, with ${\cal L}_{\rm SM}$ 
the lagrangian of the Standard Model.
This procedure is conceptually strictly related to the greek tragedy {\em Deus-ex-machina}, in which an all-mighty God or Goddess is brought to the front of the stage by a hidden crane to solve an apparently unsolvable mess by a snap of fingers. 
In this paper, I tried to translate into a mathematical form a slightly different starting point: consider the {\em brane} a defect in the extra-dimension and 
let fields that do propagate into the bulk interact with it. The origin of the defect is beyond the scope of this article, but it may be ascribed to the underlying quantum structure of the gravitational vacuum, yet to be understood, as no consistent theory of quantum gravity exists. As it is, unfortunately, the defect is still an external object introduced by hand. 
However, we can compute its effect on the fields living in the bulk by studying how their spectrum in the compact space-time is modified by the existence of the brane.
In particular, we may study if the presence of the brane in a compact space-time may induce localization of the fields (as it is found to be the case). 

I restricted myself to the simplest case of a single extra-dimension where, necessarily, the defect is represented
by a $\delta$-function  in a ${\cal M}_4 \times {\cal S}_1$ space-time (in more than one extra-dimension more complicated structures, such as vortices, could be considered). 
Then, a single, real massless scalar field $\Phi$ is coupled with the brane with a single, renormalizable interaction. 
This way to introduce the coupling of a bulk field with a $\delta$-potential is borrowed from the existing literature regarding Casimir energy computations
in presence of non-ideal boundary conditions, but it can be shown in the language of effective field theory that there is a unique relevant operator
that may couple $\Phi$ with a defect. In the realm of phenomenology, the idea was already used in the famous papers by L. Randall and R. Sundrum at the
end of the '90s. 

After formulating the model, I have solved analytically the 5D equation of motion for the field $\Phi$, finding an exact expression for the Kaluza-Klein spectrum of the model 
as a function of the coupling $q$ between field and brane. The exact solution has been found by applying the Burniston-Siewert method to solve transcendental equations,
explained in the Appendix. Applying the same method, I have also derived the exact KK-spectrum of a model in which the coupling between bulk field and brane
is achieved through an {\em irrelevant} BLKT.  Notice that the coupling $q$ is dimensionful in five dimensions and proportional to the only scale present in the model, 
the brane tension $f$. However, an adimensional coupling $\alpha$ (that can be used for perturbative expansion) emerges naturally when computing the spectrum of the model multiplying $q$ by the compactification radius $R$. I have then derived the weak and strong coupling limits of the Kaluza-Klein masses as functions of $\alpha$: 
in the weak coupling limit, the approximate results found in Ref.~\cite{Dienes:1998sb} are easily recovered; in the strong coupling limit, a perturbative expansion 
in powers of $1/\alpha$ reproduces correctly the limits $\alpha \to \infty$ found in the same paper and in the literature regarding BLKT, extending them to large but not infinite 
$\alpha$ and showing how the limits are exponentially approached. An important result is that, when a negative coupling between the scalar field and the brane is considered, 
the would-be zero-mode $\psi_{0^-}^e$ gets trapped into the brane, that behaves like a {\em crevasse}. As a consequence, the zero-mode wave-function falls exponentially in the extra-dimension and the mode is an effective, localized, four-dimensional field (exactly as it happens to the graviton zero-mode in the second paper of Randall and Sundrum 
on warped extra-dimensions). I studied, eventually, the KK-decomposition of the same action once a renormalizable cubic self-interaction term $\Phi^3 (x,y)$ is included. 
The knowledge of the exact spectrum for any value of $\alpha$ permits to derive easily the 4-dimensional couplings of KK-modes by computing their overlap
in the extra-dimension. Of particular interest is the derivation of the interaction between the localized would-be zero-mode $\psi_{0^-}^e$ and the massive
KK-modes for $\alpha < 0$. It can be shown that all couplings (but the cubic self-interaction of the 4-dimensional localized field) vanish in the limit $\alpha \to \infty$:
the impact of massive KK-modes on the phenomenology of localized modes is strongly suppressed for large $\alpha$ ({\em i.e.} in the {\em stiff brane} limit).

This is a very promising starting point to further explore the model and to see if similar phenomena occur when replacing $\Phi$ with fermion fields.

\newpage

\section*{Acknowledgements}

I am strongly indebted with many people, whose comments have been invaluable and without whom it would have been impossible for me to complete this paper.
First of all, I thank Pilar Hern\'andez. After her, in alphabetical order, I would like to thank P. Coloma, Enrique Fern\'andez Mart\'{\i}nez, Esperanza L\'opez, 
Maurizio Lusignoli, Olga Mena, Mario Paolucci, Carlos Pena, Alberto Ramos and Nuria Rius.

This work was partially supported by grants ITN INVISIBLES (Marie Curie Actions, PITN-GA-2011-289442), 
FPA2011-29678, and PROMETEOII/2014/050.

\appendix

\section{The Burniston-Siewert method}
\label{sec:appA}

In a series of papers at the beginning of the '70's, E. Burniston and C. Siewert developed a method to solve a rather broad class of
transcendental equations, whose results have been used in Sect.~\ref{sec:eigenmodes}. In this Appendix, I will shortly review the method by 
applying it to two similar transcendental equations: 
\begin{equation}
\label{eq:tanxialphaxi}
 \tan \xi  =  \alpha \xi \, , \\
\end{equation}
and 
\begin{equation}
\label{eq:xitanxialpha}
\xi \tan \xi  =  \alpha \, .
\end{equation}
The solution to eq.~(\ref{eq:tanxialphaxi}) gives the roots of eq.~(2.34) of Ref.~\cite{Dienes:1998sb}, that can be easily compared with the approximate solution presented in that paper. 
On the other hand, the solution of eq.~(\ref{eq:xitanxialpha}) gives the Kaluza-Klein spectrum presented in Sect.~\ref{sec:eigenmodes}. 

\subsection{Solving $\tan \xi = \alpha \xi$}

I will first consider the transcendental equation $f (\xi,\alpha) = 0$, with $\xi$ a real variable and $\alpha$ a real parameter, as its solution using
the Burniston-Siewert method represents a simple illustrative example. First, replace the variable $\xi$ with a complex variable $z$, 
\begin{equation}
\xi \to  \frac{i}{\alpha z}
\end{equation}
such that
\begin{equation}
\label{eq:firstsubstitutiontanxialphaxi}
\tan \xi = \alpha \xi \longrightarrow \tan \left ( \pm n \pi + \frac{i}{\alpha z} \right ) = \frac{i}{z} \qquad n = 0,  1, \dots
\end{equation}
where the periodicity of the tangent on the real axis is replaced by its multi-valuedness in the complex plane, with the integer $n$ labelling the different Riemann sheets of the function. Eq.~(\ref{eq:firstsubstitutiontanxialphaxi}) can be inverted to get: 
\begin{equation}
\label{eq:inverting}
\log \frac{(z +1)}{(z -1)} = \mp \, 2 \pi n i \alpha z + \frac{2}{\alpha z} \, ,
\end{equation}
where "$\log$" is the principal value of the logarithm in the complex plane, with branching points at $z = \pm 1$ and a branch cut going from $-1$ to $1$ on the real axis. 
Eventually, we can write this equation as: 
\begin{equation}
\label{eq:tobesolved1}
\Lambda (z) = 1 + \frac{\alpha z}{2} \left [ \log \frac{(z - 1)}{(z + 1)} \pm \, 2 \pi n i \right ] = 0 \, .
\end{equation}
Finding the roots $z_0$ of $\Lambda (z)$ will give, then, the solutions of eq.~(\ref{eq:tanxialphaxi}) after the inverse substitution $z_0 \to i / \alpha \xi_0$. 

Solving $\Lambda (z) = 0$ can be as hard as to solve the original equation. However, the explicit form of functions that are analytic on the complex plane with the exception of a boundary can be determined using what is called a Riemann-Hilbert (RH) boundary value problem (see, for example, Ref.~\cite{ablowitz2003complex}).  First, write $\Lambda (z)$ conveniently: 
\begin{equation}
\left \{
\begin{array}{lll}
\Lambda_0 (z) & = & 1 + \frac{\alpha z}{2} \log \frac{(z-1)}{(z+1)} \\
\\
\Lambda_n (z) & = & \Lambda_0 (z) + \pi n i \alpha z
\end{array}
\right .
\end{equation}
Notice that $\Lambda_0 (z)$ is symmetric for $z \to - z$ and that, at infinity, $\Lambda_0 (z \to \infty) = 1 - \alpha$.
Consider first the case $n = 0$, for which $\Lambda (z) = \Lambda_0 (z)$, and write the scalar homogeneous Riemann-Hilbert boundary value problem
\begin{equation}
\label{eq:RHproblem1}
X_0^+ (t) = g_0 (t) \, X_0^- (t) \qquad {\rm for} \; t \in {\cal R} \; {\rm and} \; 0 < t < 1 \, 
\end{equation}
where
\begin{equation}
g_0 (t) = \frac{\Lambda_0^+ (t)}{\Lambda_0^- (t)}
\end{equation}
and $\Lambda_0^\pm (t)$ is the limiting value of the function $\Lambda_0 (z)$ when approaching the real axis $z \to t$ from above or below, respectively: 
\begin{equation}
\Lambda_0^\pm (t) = \lambda (t) \pm i \pi \frac{\alpha t}{2}
\end{equation}
with
\begin{equation}
\lambda (t) = 1 + \frac{\alpha t}{2} \ln \frac{(1-t)}{(1+t)} \, ,
\end{equation}
being "$\ln$" the standard natural logarithm. Notice that, as $\Lambda_0 (z)$ is symmetric for $z \to - z$, we can restrict the RH problem to the interval $0 < t < 1$, only. 
The function $g_0 (t)$, for the particular case at hand, can be written as: 
\begin{equation}
g_0 (t) = \exp \left [ 2 i \arg \Lambda_0^+ (t) \right ] \, ,
\end{equation}
since $|\Lambda_0^+(t)| = |\Lambda_0^-(t)|$ and, by symmetry, $\arg \Lambda_0^- (t) = - \arg \Lambda_0^+ (t)$. Notice that $\arg \Lambda_0^+ (t)$
is assumed to be the principal value of the argument\footnote{For $\alpha > 0$ it always exists a $\bar t \in ]0,1[$ for which $\lambda (\bar t) = 0$. At that point, 
$\arg \Lambda_0^+ (\bar t) \to \pm \pi/2$ depending if $(t - \bar t ) \to 0^-$ or $=0^+$, respectively. As we take the principal value of the argument of $\Lambda_0^+ (t)$, 
for $t \in ] 0, \bar t]$ we have $\arg \Lambda_0^+ (t) = \arctan \left [ (\pi \alpha t / 2) / \lambda (t) \right ]$, whereas for $t \in ] \bar t, 1[$ we have 
$\arg \Lambda_0^+ (t) = \arctan \left [ (\pi \alpha t / 2) / \lambda (t) \right ] + \pi$, thus restoring the continuity of $g_0 (t)$ along the interval $t \in ] 0, 1[$.} .

A so-called {\em canonical solution} to the RH problem is given \cite{muskhelishvili1953singular} by: 
\begin{equation}
X_0 (z) = (z - a)^\lambda \times (z-b)^\mu \times \exp \Gamma_0 (z) \, , 
\end{equation}
with $\lambda, \mu$ integers, and $a,b$ the ending points of the interval on which eq.~(\ref{eq:RHproblem1}) is evaluated (in our case, $a = 0$ and $b = 1$). 
The function $\Gamma_0 (z)$ is given by:
\begin{equation}
\Gamma_0 (z) = \frac{1}{2 \pi i} \int_a^b d \tau \frac{1}{\tau - z} \log g_0 (\tau)  \, .
\end{equation}
The degree of divergence of $\Gamma_0 (z)$ at the ending points is given by:
\begin{equation}
\label{eq:divergenceGamma}
\Gamma_0 (z) \rightarrow \left \{ 
\begin{array}{l}
- \frac{1}{2 \pi i} \log g_0 (a) \log (z-a) + \Gamma_a (z)  = \gamma_a  \log (z-a) + \Gamma_a (z)  \qquad {\rm for} \;  z \to a \\
\\
 \frac{1}{2 \pi i} \log g_0 (b) \log (z-b) + \Gamma_b (z)  = \gamma_b \log (z-b) + \Gamma_b (z)   \qquad {\rm for} \;  z \to b \\
\end{array}
\right .
\end{equation}
where $\Gamma_a (z)$ and $\Gamma_b (z)$ are functions that goes to a finite value for $z \to a, b$, respectively. The degree of $X_0 (z)$ when approaching
the endings $a,b$ is, therefore:
\begin{equation}
X_0 (z) \to \left \{
\begin{array}{l}
(z-a)^{\lambda + \Re [\gamma_a]} \qquad {\rm for} \;  z \to a \, , \\
\\
(z-b)^{\mu + \Re [\gamma_b]} \qquad {\rm for} \;  z \to b \, .
\end{array}
\right .
\end{equation}
A necessary condition for $X_0 (z)$ to be integrable at both ending points is that $-1 < \lambda + \Re [\gamma_a] < 1$ and $-1 < \mu + \Re [\gamma_b]  < 1$.
As in the case at hand $\Re [\gamma_a] = \gamma_a = 0$ and $\Re [\gamma_b] = \gamma_b = 1$ then\footnote{Since both $\Re [\gamma_a]$ and $\Re [\gamma_b]$
are integers, the endings of the interval $t \in ] 0, 1[$ are called {\em special endings}.}, trivially, we have $\lambda = 0, \mu = -1$. 
Therefore, in our case a canonical solution is:
\begin{equation}
X_0 (z) = \frac{1}{z-1} \times \exp \Gamma_0 (z) \, .
\end{equation}
We can construct another solution to the RH problem valid for the whole interval $-1 < t < 1$. This solution is given by:
\begin{equation}
\psi (z) = \frac{\Lambda_0 (z)}{X_0 (-z)}
\end{equation}
For $z \to t$ and $ 0 < t < 1$, $\psi (z) \to \psi^\pm (t)$. It is easy to show that $\psi^\pm (t)$ is a trivial solution to eq.~(\ref{eq:RHproblem1}) since $X_0 (-z)$ 
has no singularities for $z \to t$ and, therefore, $X_0^+ (-t) = X_0^-(-t)$ for $0 < t < 1$. On the other hand, $\psi (z)$ is continuous over the whole interval $-1 < t < 1$, since
$\Lambda_0 (-z) = \Lambda_0 (z)$ and $1/X_0 (-z)$ is non-singular everywhere.
A general solution to the scalar homogeneous RH problem is given by $ \psi (z) = X_0 (z) P_m (z)$, with $P_m (z)$ a polynomial of degree $m$. Therefore, 
\begin{equation}
\psi (z) = \frac{\Lambda_0 (z)}{X_0 (-z)} = X_0 (z) P_m (z)
\end{equation}
and
\begin{equation}
\label{eq:solutiontanxialphaxi1}
\Lambda_0 (z) = X_0 (z) X_0 (-z) P_m (z)
\end{equation}
We can now use our knowledge of $\Lambda_0 (z)$ to fix some of the arbitrariness of $P_m (z)$. First of all, notice that $\Lambda_0 (z) \to (1-\alpha)$ for
$z \to \infty$. Since $X_0 (z) X_0 (-z)$ is $ {\cal O} (z^{-2})$ for $z \to \infty$, the polynomial must be ${\cal O} (z^2)$ ({\em i.e. $m = 2$})
\begin{equation}
P_2 (z) = A_0 + A_1 z + (1-\alpha) z^2 \, .
\end{equation}
Being $\Lambda_0 (z)$ symmetric under $z \to - z$, $A_1 = 0$. Rescaling conveniently $A_0 =  (1 - \alpha) \, z_0^2$,  we can write:
\begin{equation}
\label{eq:solutiontanxialphaxi2}
\Lambda_0 (z) = - X_0 (z) X_0 (-z) (1- \alpha) \left ( z^2 - z_0^2 \right ) \, .
\end{equation}
Since $X_0 (z)$ does not vanish on the complex plane, the roots of $\Lambda_0 (z)$ are given by the roots of the polynomial 
$P_2(z)$. It is now trivial to compute them by putting $z = 0$ in eq.~(\ref{eq:solutiontanxialphaxi2}):
\begin{equation}
z_0 = i \,  (\alpha - 1)^{-1/2} \, X_0 (0)^{-1} \qquad  {\rm for } \qquad \alpha > 0  \, ,
\end{equation}
whereas there is no solution for $\alpha < 0$. For $\alpha \in [0, 1]$ we find two imaginary roots at $z = \pm z_0$ (that become degenerate for $\alpha = 1$), 
whilst there are two real roots for $\alpha > 1$.  Eventually, 
\begin{equation}
\xi_0 = \pm \frac{1}{\alpha} (\alpha - 1)^{1/2} \exp \left [ \frac{1}{\pi} \int_0^1 dt \frac{1}{t} \arg \Lambda_0^+ (t) \right ] \qquad (\alpha > 0) \, ,
\end{equation}
with $ \arg \Lambda_0^+(t)$ being the principal value of the argument of $\Lambda_0^+ (t)$. Remind that there is no solution to $\Lambda_0 (z) = 0$ for $\alpha < 0$.

We must now solve eq.~(\ref{eq:tobesolved1}) for $n \geq 1$. Notice first that $\Lambda_n (-z) = \Lambda_{-n} (z)$.  To symmetrize $\Lambda_n (z)$, thus, we define:
\begin{equation}
\Omega_n (z) = \Lambda_n (z) \Lambda_{-n} (z) = \Lambda_n (z) \Lambda_n (- z)
\end{equation}
and solve the scalar homogeneous RH boundary value problem
\begin{equation}
\label{eq:RHproblem2}
X_n^+ (t) = g_n (t) \, X_n^- (t)  \qquad {\rm for} \; t \in {\cal R} \; {\rm and} \; 0 < t < 1 \, 
\end{equation}
where
\begin{equation}
g_n (t) = \frac{\Omega_n^+ (t)}{\Omega_n^- (t)} = \exp \left [ 2 i \arg \Omega_n^+ (t) \right ] \, ,
\end{equation}
and
\begin{equation}
\Omega_n^+ (t) = \left [ \Lambda_0^+ (t) \right ]^2 + \pi^2 \alpha^2 n^2 t^2 \, .
\end{equation}
In this case, it is easy to see that $\arg \Omega_n^+ (t)$ is a smooth continuous function for $t \in ] 0, 1[$. 
The canonical solution to eq.~(\ref{eq:RHproblem2}) is: 
\begin{equation}
X_n (z) = \exp \Gamma_n (z) \, , 
\end{equation}
with
\begin{equation}
\Gamma_n (z) = \frac{1}{\pi } \int_0^1 d \tau \frac{1}{\tau - z} \arg \Omega_n^+ (\tau)  \, .
\end{equation}
Notice that, in this case, the exponents $\lambda, \mu$ computed at the endings $t = 0, 1$ are trivially found to vanish. As before, we can relate the 
complex function $\Omega_n (z)$ with a general solution to eq.~(\ref{eq:RHproblem2}) to find:
\begin{equation}
\label{eq:solutiontanxialphaxin}
\Omega_n (z) = - X_n (z) X_n (-z) \left ( z^2 - z_n^2 \right ) \pi^2 \alpha^2 n^2 \, ,
\end{equation}
with roots given by:
\begin{equation}
z_n = i \,  (\pi \alpha n)^{-1} \, X_n (0)^{-1} \qquad  {\rm for } \qquad \alpha \in ] - \infty, + \infty [ \, ,
\end{equation}
and, eventually,
\begin{equation}
\xi_n = \pm \pi n \exp \left [ \frac{1}{\pi} \int_0^1 dt \frac{1}{t} \arg \Omega_n^+ (t) \right ] \qquad n = 1, 2, \dots,\; \alpha \in ] - \infty, + \infty [
\end{equation}

\subsection{Solving $\xi \tan \xi = \alpha$}

Consider the transcendental equation $f (\xi,\alpha) = 0$ with $\xi$ a real variable and $\alpha$ a real parameter. 
First, replace the variable $\xi$ with the complex variable $z$:
\begin{equation}
\xi \to i \alpha z \, ,
\end{equation}
such that
\begin{equation}
\label{eq:firstsubstitution}
\xi \tan \xi = \alpha \xi \longrightarrow \tan \left ( \pm n \pi + i \alpha z \right ) = - \frac{i}{z} \qquad n = 0, 1, \dots
\end{equation}
Inverting this equation, we get: 
\begin{equation}
1 - \frac{1}{2 \alpha z} \left [ \log \frac{(z-1)}{z+1} \pm 2 \pi n i  \right ]  = 0 \, .
\end{equation}
We can safely multiply by $z$, as $z = 0$ is a solution of eq.~(\ref{eq:firstsubstitution}) only in the trivial case $\alpha = 0$. Therefore, the equation to be solved is:
\begin{equation}
\label{eq:tobesolved3}
\Lambda (z) = z - \frac{1}{2 \alpha} \left [ \log \frac{(z-1)}{z+1} \pm 2 \pi n i  \right ]  = 0 \, .
\end{equation}
Notice that $\Lambda (z)$ is antisymmetric in $z$. A necessary condition to use the Burniston-Siewert method is, on the other hand, that $\Lambda (z)$ be symmetric in $z$.
Let's define, therefore:
\begin{equation}
\left \{
\begin{array}{lll}
\Lambda_0 (z) & = & z \left [ z - \frac{1}{2 \alpha} \log \frac{(z-1)}{(z+1)} \right ] \\
\\
 \Lambda_n (z) & = & \Lambda_0 (z) - i  \frac{\pi n z}{\alpha} \qquad n = \pm 1, \pm 2, \dots
\end{array}
\right .
\end{equation}
and
\begin{equation}
\Omega_n (z) = \Lambda_n (z) \Lambda_{-n} (z) \, ,
\end{equation}
such that both $\Lambda_0 (z)$ and $\Omega_n (z)$ are symmetric for $z \to - z$. For $n = 0$, we must solve the scalar homogeneous RH boundary value problem:
\begin{equation}
\label{eq:RHproblem3}
X_0^+ (t) = g_0 (t) \, X_0^- (t)  \qquad {\rm for} \; t \in {\cal R} \; {\rm and} \; 0 < t < 1 \, .
\end{equation}
The function $g_0 (t)$ is given by:
\begin{equation}
g_0 (t) = \frac{\Lambda_0^+ (t)}{\Lambda_0^-(t)} \, ,
\end{equation}
with
\begin{equation}
\Lambda_0^\pm (t) = \lambda (t) \mp i \frac{\pi t}{2 \alpha}
\end{equation}
and
\begin{equation}
\lambda (t) = t \left [ t - \frac{1}{2 \alpha}  \ln \frac{(1-t)}{(1+t)} \right ] \, .
\end{equation}
The canonical solution to eq.~(\ref{eq:RHproblem3}) is given by:
\begin{equation}
X_0 (z) = z^\lambda (z-1)^\mu \times \exp \Gamma_0 (z) \, ,
\end{equation}
where
\begin{equation}
\Gamma_0 (z) = \frac{1}{\pi} \int_0^1 dt \frac{1}{t - z} \arg \Lambda_0^+ (t) \, .
 \end{equation}
The computation of the exponents $\lambda, \mu$ is more involved than in the previous case. Also in this case $\gamma_a, \gamma_b$ are real; however,
\begin{equation}
\left \{
\begin{array}{l}
\gamma_a= \left \{ 
\begin{array}{l}
1/2 \qquad {\rm for} \; \alpha > 0 \\
\\
-1/2 \qquad {\rm for} \; \alpha < 0
\end{array}
\right .
\; \longrightarrow \qquad
\lambda +
\left \{ 
\begin{array}{l}
\lfloor 1/2 \rfloor = -1 \; {\rm or} \; 0 \qquad {\rm for} \; \alpha > 0 \\
\\
 \lfloor -1/2 \rfloor = -1 \; {\rm or } \; 0 \qquad {\rm for} \; \alpha < 0
\end{array}
\right .
\\
\\
\gamma_b = \left \{ 
\begin{array}{l}
0 \qquad {\rm for} \; \alpha > 0 \\
\\
1 \qquad {\rm for} \; \alpha < 0
\end{array}
\right .
\qquad \longrightarrow \qquad
\mu = \left \{ 
\begin{array}{l}
0 \qquad {\rm for} \; \alpha > 0 \\
\\
-1 \qquad {\rm for} \; \alpha < 0
\end{array}
\right .
\end{array}
\right .
\end{equation}
We can see that, in this case, there are two possible choices for the exponent $\lambda$ that fulfill the requirement $\left | \lambda + \Re [\gamma_a] \right | < 1$. This 
is a standard result when the ending is not a {\em special ending}, as it is the case for $t \to 0$. Consider first the case $\alpha > 0$. 
In this case, we have either
\begin{equation}
(1) \qquad X_0 (z) = \exp \Gamma_0 (z) \qquad {\rm or} \qquad (2) \qquad  X_0 (z) = \frac{1}{z} \exp \Gamma_0 (z) \, .
\end{equation}
In case (1) we have: 
\begin{equation}
\Lambda_0 (z) = \exp \left [ \Gamma_0 (z) + \Gamma_0 (-z) \right ] \times P_m (z)
\end{equation}
with the constraint that
\begin{equation}
\lim_{z \to \infty} \Lambda_0 (z) \to z^2 = A_m z^m \longrightarrow \left \{ 
\begin{array}{l}
m = 2 \\
A_m = 1
\end{array}
\right .
\end{equation}
and, thus, 
\begin{equation}
\label{eq:solutiontxitanxialpha}
\Lambda_0 (z) = \exp \left [ \Gamma_0 (z) + \Gamma_0 (-z) \right ] \times \left ( A_0 + z^2 \right ) \, .
\end{equation}
To find the zeroes of $\Lambda_0 (z)$ we must look for the zeroes of the polynomial $P_2 (z)$, {\em i.e.}  $z = \pm \sqrt{- A_0}$. In order to fix the coefficient $A_0$ of 
the polynomial, we take the limit for $z \to 0$ on both sides: 
\begin{equation}
\label{eq:findingroots}
A_0 = - z_0^2 = \lim_{z \to 0} \exp \left [ \Gamma_0 (z) + \Gamma_0 (-z) \right ] \Lambda_0 (z) \, .
\end{equation}
Notice that $\lim_{z \to 0} \Lambda_0 (z) = 0$. However, we know that $\Gamma_0 (z)$ diverges for $z \to 0$ as $z^{1/2}$ (for $\alpha > 0$). Therefore, the limit of the ratio 
of left and right hand side of eq.~(\ref{eq:xitanxialpha}) is a constant. Using eq.~(\ref{eq:divergenceGamma}), we define:
\begin{eqnarray}
\Gamma_0 (z) &=& \frac{1}{\pi} \int_0^1 dt \frac{1}{t - z} \arg \Lambda_0^+ (t) 
+ \frac{1}{\pi} \int_0^1 dt \frac{\arg \Lambda_0^+ (0) }{t - z} - \frac{1}{\pi} \int_0^1 dt \frac{\arg \Lambda_0^+ (0) }{t - z}  \nonumber \\
\\
&=& \frac{1}{\pi} \int_0^1 dt \frac{\arg \Lambda_0^+ (0) }{t - z}  + \, \Gamma_a (z) \, , \nonumber
\end{eqnarray}
where 
\begin{equation}
\Gamma_a (z) = \frac{1}{\pi} \int_0^1 dt \frac{1}{t-z} \left [ \arg \Lambda_0^+ (t) - \arg \Lambda_0^+ (0) \right ]
\end{equation}
 is finite for $z \to 0$. Therefore, 
\begin{eqnarray}
z_0^2 &=& - \exp \left [ - 2 \Gamma_a (0)  \right ] \times \lim_{z \to 0} \frac{1}{\sqrt{z}} \frac{1}{\sqrt{-z}} \Lambda_0 (z) 
        =  - \exp \left [ - 2 \Gamma_a (0)  \right ] \times \lim_{z \to 0} \left ( \frac{1}{i z} \right ) \left ( i \frac{\pi z}{2 \alpha }\right ) \nonumber \\
       & = & - \frac{\pi}{2 \alpha } \times  \exp \left [ - 2 \Gamma_a (0)  \right ] \, .
\end{eqnarray}
Notice that we took into account that, according to the definition of principal value of the complex logarithm, $\log[ (z-1)/(z+1)] \to - i \pi $ for $z \to 0$.
Eventually, 
\begin{equation}
\label{eq:resultxitanxialpha}
z_0 = i \sqrt{\frac{\pi}{2 \alpha}} \exp \left \{ -\frac{1}{\pi} \int_0^1 dt \frac{1}{t} \left [ \arg \Lambda_0^+ (t) + \frac{\pi}{2} \right ]
\right \} \qquad {\rm for} \qquad \alpha > 0
\end{equation}
In case (2) we have: 
\begin{equation}
\Lambda_0 (z) = - \frac{1}{z^2 }\exp \left [ \Gamma_0 (z) + \Gamma_0 (-z) \right ] \times P_m (z)
\end{equation}
with the constraint that
\begin{equation}
\lim_{z \to \infty} \Lambda_0 (z) \to z^2 = - \frac{1}{z^2} A_m z^m \longrightarrow \left \{ 
\begin{array}{l}
m = 4 \\
A_m = -1
\end{array}
\right .
\end{equation}
Since for $z \to 0$ we have $\Lambda_0 (z) \to 0$ and $\lim_{z \to 0} \Gamma_0 (z) \propto z^{1/2}$, we have that necessarily $A_0 = 0$. Therefore,
\begin{equation}
\label{eq:solutiontxitanxialpha2}
\Lambda_0 (z) = - \exp \left [ \Gamma_0 (z) + \Gamma_0 (-z) \right ] \times \left ( A_2 - z^2 \right ) \, ,
\end{equation}
that gives the same result as before with the choice $A_2 = z_0^2$.  It is easy to show that, using the same rules as before, also in the case of negative coupling $\alpha < 0$
we get the same result of eq.~(\ref{eq:resultxitanxialpha}), but for a change in the sign of the $\pi/2$ term in the definition of $\Gamma_a (z)$. Eventually:
\begin{equation}
\label{eq:resultxitanxialpha}
z_0 = i \sqrt{\frac{\pi}{2 \alpha}} \exp \left \{ -\frac{1}{\pi} \int_0^1 dt \frac{1}{t} \left [ \arg \Lambda_0^+ (t) + \frac{\pi}{2} {\rm sign} (\alpha )\right ]
\right \} \qquad {\rm for \;  any} \; \alpha 
\end{equation}

For $n \neq 0$, the RH problem to be solved is: 
\begin{equation}
\label{eq:RHproblem4}
X_n^+ (t) = g_n (t) \, X_n^- (t)  \qquad {\rm for} \; t \in {\cal R} \; {\rm and} \; 0 < t < 1 \, ,
\end{equation}
where the function $g_n (t)$ is given by:
\begin{equation}
g_n (t) = \frac{\Omega_n^+ (t)}{\Omega_n^-(t)} \, ,
\end{equation}
with
\begin{equation}
\Omega_n^\pm (t) = \left [ \Lambda_0^\pm (t) \right ]^2 + \frac{\pi^2 n^2 t^2}{\alpha^2}  = 
\left \{ \lambda^2 (t) + \left ( n^2 - \frac{1}{4} \right ) \frac{\pi^2 t^2}{\alpha^2} \right \} \mp i \frac{\pi t }{\alpha} \lambda (t) \, .
\end{equation}
Also in this case $\arg \Omega_n^- (t) = - \arg \Omega_n^+ (t)$ for any value of $\alpha$ and $n$. Thus, a canonical solution to eq.~(\ref{eq:RHproblem4}) is:
\begin{equation}
X_n (z) = z^\lambda (z-1)^\mu \times \exp \Gamma_n (z) \, ,
\end{equation}
where
\begin{equation}
\Gamma_n (z) = \frac{1}{\pi} \int_0^1 dt \frac{1}{t - z} \arg \Omega_n^+ (t) \, .
 \end{equation}
The exponents $\lambda,\mu$ are trivially found to vanish, in this case. Thus, the ending points $(a,b) = (0,1)$ are {\em special endings} and $\Gamma_n (z)$ is finite there.
Eventually,
\begin{equation}
\Omega_n (z) = \exp \left [ \Gamma_n (z) + \Gamma_n (-z) \right ] P_m (z)
\end{equation}
and, since $\Omega_n (z) \to z^4$ for $z \to \infty$, we have $m = 4$ and $A_m = 1$. Taking the limit $z \to 0$, we find easily that $A_0 = 0$. Therefore, 
\begin{equation}
\Omega_n (z) = z^2 \exp \left [ \Gamma_n (z) + \Gamma_n (-z) \right ] (z^2 - z_n^2) \, .
\end{equation}
The roots of $\Omega_n (z)$ can be found as in eq.~(\ref{eq:findingroots}),
\begin{equation}
z_n^2 = - \exp \left [ - 2 \Gamma_n (0)  \right ] \times \lim_{z \to 0} \frac{1}{z^2} \Omega_n (z) 
        =  -  \frac{\pi^2}{\alpha^2} \left (n^2 - \frac{1}{4} \right ) \times \exp \left [ - 2 \Gamma_n (0)  \right ]
 \end{equation}

After the inverse substitution $z \to - i \xi/\alpha$, we eventually get the roots of the transcendental equation $\xi \tan \xi = \alpha$:
\begin{equation}
\left \{
\begin{array}{lll}
\xi_0 & = & \pm \sqrt{\frac{\pi \alpha}{2}} \, \exp \left \{ - \frac{1}{\pi}  \int_0^1 dt \frac{1}{t}  \left [ \arg \Lambda_0^+ (t) + \frac{\pi}{2} \, {\rm sign} (\alpha) \right ] \right \} \, , \\
\\
\xi_n &=& \pm \frac{\pi}{2} \sqrt{4 n^2 -1}  \, \exp \left \{ - \frac{1}{\pi}  \int_0^1 dt \frac{1}{t}  \arg \Omega_n^+ (t) \right \}  \, ,
\end{array}
\right .
\end{equation}
{\em i.e.} the results shown in Sect.~\ref{sec:evenmodes}, eq.~(\ref{eq:xitanxialphasolutions}).

\bibliographystyle{h-elsevier}
\bibliography{refbrane}

\end{document}